\documentclass[sigconf]{acmart}

\copyrightyear{2025}
\acmYear{2025}
\setcopyright{cc}
\setcctype{by-sa}
\acmConference[SC Workshops '25]{Workshops of the International Conference for High Performance Computing, Networking, Storage and Analysis}{November 16--21, 2025}{St Louis, MO, USA}
\acmBooktitle{Workshops of the International Conference for High Performance Computing, Networking, Storage and Analysis (SC Workshops '25), November 16--21, 2025, St Louis, MO, USA}
\acmDOI{10.1145/3731599.3767508}
\acmISBN{979-8-4007-1871-7/2025/11}

\usepackage{algorithmic}
\usepackage{algorithm}
\usepackage{graphicx}
\usepackage{url}
\usepackage{textcomp}
\usepackage{xcolor}
\usepackage{mathtools}
\usepackage{sc25repro}
\def\BibTeX{{\rm B\kern-.05em{\sc i\kern-.025em b}\kern-.08em
    T\kern-.1667em\lower.7ex\hbox{E}\kern-.125emX}}

\newcommand{\figref}[1]{Fig.~\ref{#1}}

\makeatletter
\DeclareRobustCommand{\change}{%
  \@bsphack
  \leavevmode
  \@esphack
}
\DeclareRobustCommand{\stopchange}{%
  \@bsphack
  \@esphack
}
\makeatother

\graphicspath{{./img/}}
\DeclareGraphicsExtensions{.pdf,.png,.jpg}

\newcommand{\PDCKTH}{
\affiliation{%
  \department{PDC Center for High Performance Computing}
  \institution{KTH Royal Institute of Technology}
  \city{Stockholm}
  \country{Sweden}}
}
\newcommand{\NVIDIA}{
\affiliation{%
  \institution{NVIDIA}
  \city{Santa Clara}
  \state{CA}
  \country{USA}}
}

\begin{document}

\title{Redesigning GROMACS Halo Exchange: Improving Strong Scaling with GPU-initiated NVSHMEM}

\author{Mahesh Doijade}
\NVIDIA
\email{mdoijade@nvidia.com}
\orcid{0009-0003-5953-0436}

\author{Andrey Alekseenko}
\PDCKTH
\email{andreyal@kth.se}
\orcid{0000-0003-4906-7241}

\author{Ania Brown}
\NVIDIA
\email{anbrown@nvidia.com}
\orcid{0009-0000-7116-7535}

\author{Alan Gray}
\NVIDIA
\email{alang@nvidia.com}
\orcid{0009-0009-7731-1855}

\author{Szil\'ard P\'all}
\PDCKTH
\email{pszilard@kth.se}
\orcid{0000-0003-0603-5514}


\begin{abstract}

Improving time-to-solution in molecular dynamics simulations often requires strong scaling due to fixed-sized problems.
GROMACS is highly latency-sensitive, with peak iteration rates in the sub-millisecond, making scalability on heterogeneous supercomputers challenging.
MPI's CPU-centric nature introduces additional latencies on GPU-resident applications' critical path, hindering GPU utilization and scalability.
To address these limitations, we present an NVSHMEM-based GPU kernel-initiated 
redesign of the GROMACS domain decomposition halo-exchange algorithm.
Highly tuned GPU kernels fuse data packing and communication, leveraging hardware latency-hiding for fine-grained overlap.
We employ kernel fusion across overlapped data forwarding communication phases and utilize the asynchronous copy engine over NVLink to optimize latency and bandwidth. Our GPU-resident formulation greatly increases communication-computation overlap, improving GROMACS strong scaling performance across NVLink by up to 1.5x (intra-node) and 2x (multi-node), and up to 1.3x  multi-node over NVLink+InfiniBand. 
This demonstrates the profound benefits of GPU-initiated communication for strong-scaling a broad range of latency-sensitive applications.


\end{abstract}



\maketitle
\section{Introduction}

Molecular dynamics (MD) simulations are a critical tool in domains like computational chemistry, biophysics, biochemistry, and materials science and are key to complement laboratory experiments: they are among the top cycle consuming HPC simulation workloads. Therefore, improving performance, scalability, and efficiency has long been pursued by MD applications.

The MD field was among the early pioneers of GPU computing with great success in accelerating simulations. While some production simulations, especially of smaller molecular systems, have become routinely feasible on a single GPU, the timescale demands of MD as well as the gradual (although moderate) increase in system sizes of interest continue to require strong scaling for better time-to-solution. 

GROMACS has long demonstrated excellent strong-{}sca\-ling on commodity HPC hardware \cite{Pronk2013,Pall2014,Loeffler2012}, and continues to excel in both scaling and absolute performance on a wide range of architectures \cite{kutzner_scaling_2025}.
However, modern heterogeneous architectures and programming models pose significant scalability challenges for the latency-sensitive MD algorithms. Continued hardware improvements combined with long-term investment into optimizing the GROMACS heterogeneous parallelization have pushed production simulation time-steps into the sub-millisecond regime, with 100-200 µs at peak across multiple GPUs, where latency-hiding, asynchronous execution, and overlapping of host and device-side computation and communication are all key. 
Our previous work began to address these issues with an event-driven multi-GPU communication design which, using GROMACS' built-in thread-MPI library, implements asynchronous scheduling of all compute and communication (using DMA copies) on GPU queues. With its streamlined control path and ability to overlap launch latencies, this setup can outperform GPU-aware MPI in scaling regimes where local computation is insufficient to fully overlap communication.
 This regime is increasingly common\footnote{Depending on the size of GPU and decomposition corresponding to $<$100000-200000 atoms/GPU.}, as larger and more parallel GPUs speed up local computation, thereby exposing communication bottlenecks across a wider range of problem sizes.

While stream-aware MPI with CPU-initiated and GPU-triggered communication \cite{namashivayam_exploring_2023,bridges_understanding_2024} could replicate the benefits of our event-driven asynchronous approach, such coarse-grained approaches have not been proven in practice, and struggle to exploit NVLink’s fine-grained capability or the GPU's latency-hiding capabilities. In contrast, PGAS-style communication approaches, when combined with GPU kernel-initiated capabilities, offer the potential to do both.

This work has the following main contributions:
\begin{itemize}
    \item 
A novel, GPU-initiated formulation of the GROMACS neutral-territory domain decomposition halo exchange algorithm using NVSHMEM. This approach combines the pack/unpack and data transfers in each communication stage, and all communication stages across decomposition dimensions are fused maximizing communication concurrency by minimizing dependencies between stages through the separate handling of dependent and independent data.
\item
Optimized hierarchical synchronization that allows receiver notification to be fused with data transfer kernels.
\item 
 Optimizations for both inter- and intra-node communication with adaptive strategy based on interconnect type (NVLink vs InfiniBand) and tensor memory accelerator (TMA) async copy engine-based data movement for latency hiding.
\item

Strong scaling studies on the NVIDIA EOS cluster and a detailed critical path and overlap analysis using GPU cycle timers.
\end{itemize}

\section{Background}

\subsection{Molecular dynamics}
In classical molecular dynamics, particles are assigned positions and velocities. The calculation of forces, via a model-physics force-field and required to update positions and velocities by integrating Newton’s equations of motion, are typically the most computationally expensive component: in particular  
the non-bonded pair interactions that model van der Waals forces and Coulomb’s law. Repeated numerical integration generates samples from the model's thermodynamic ensemble, enabling observations in computational experiments.
The high cost of force evaluation, the short (typically femtosecond) time-steps and the requirement to run very long simulations to resolve real-world timescales (such as milliseconds in biological systems) combine to pose a major computational challenge. 
While other fields leverage larger problems or higher resolutions to expose greater parallelism to target larger HPC machines, most MD simulations are limited by atomistic resolution and the size of modeled systems. MD studies of biomolecules commonly simulate tens to hundreds of thousands of atoms, with simulations of millions to tens of millions of atoms becoming increasingly common. 
Reducing the wall-time per time-step remains the primary way to improve performance: strong scaling is necessary. This is routinely combined with ensemble scaling, but a faster iteration rate remains critical for reducing time-to-solution. In pursuit of faster time-steps, MD was an early adopter of GPUs, successfully using both data-center and consumer devices. This was possible because MD simulations are arithmetically intensive and have long relied on single-precision floating-point arithmetic \cite{lindahl_gromacs_2001,Harvey2009a,Friedrichs2009}.

\subsection{GROMACS}
GROMACS is a free, open source molecular dynamics community software package and one of the most widely used HPC simulation packages. Written primarily in C++17 with a wide range of backends (including CPU/SIMD, CUDA, SYCL, OpenCL, and HIP) it is highly portable across modern hardware. Combined with state-of-the-art parallel algorithms plus a bottom-up performance-oriented design, it runs efficiently from laptops to workstations to the largest supercomputers, making it a popular choice for academia, industry, and education. Most production simulations exploit a mixed precision mode which predominantly uses single-precision arithmetic\cite{Hess2008,Pall2014}.

\subsubsection*{GROMACS parallelization}

The GROMACS MD engine relies on a multi-level data and task parallelization to individually target multiple hardware levels. At the lowest level, SIMD and SIMT/GPU algorithms are implemented in highly tuned compute kernels\cite{Pall2013}. Cache- and NUMA-optimized multicore algorithms use OpenMP multithreading on CPUs. A task-parallel heterogeneous scheme utilizes event-driven multi-queue GPU schedules. Exploiting the data-parallelism in force computation, GROMACS can use both CPU and GPU concurrently, a key ingredient to its flexibility and extensibility\cite{Pall2020}. CUDA and SYCL graph support is available for GPU-resident parallelization setups intra-node.
For distributed-memory parallelization, GROMACS uses a neutral-territory domain decomposition (DD) implemented with MPI and NVSHMEM. Multiple-program multiple-data (MPMD) rank specialization is used to dedicate a subset of ranks to computing long-range interactions with the 3D FFT-based particle-mesh Ewald (PME) algorithm\cite{Hess2008} while the rest compute particle-particle interactions and carry out integration (PP ranks). Rank specialization is critical to strong scaling PME which relies on multiple 3D FFT transforms. In addition to regular MPI support, GROMACS comes with a built-in thread-MPI library based on pthreads \cite{Pronk2013}, originally designed to simplify use on single-node multicore systems without requiring an external MPI.

\begin{figure}
    \centering
    \includegraphics[width=\linewidth]{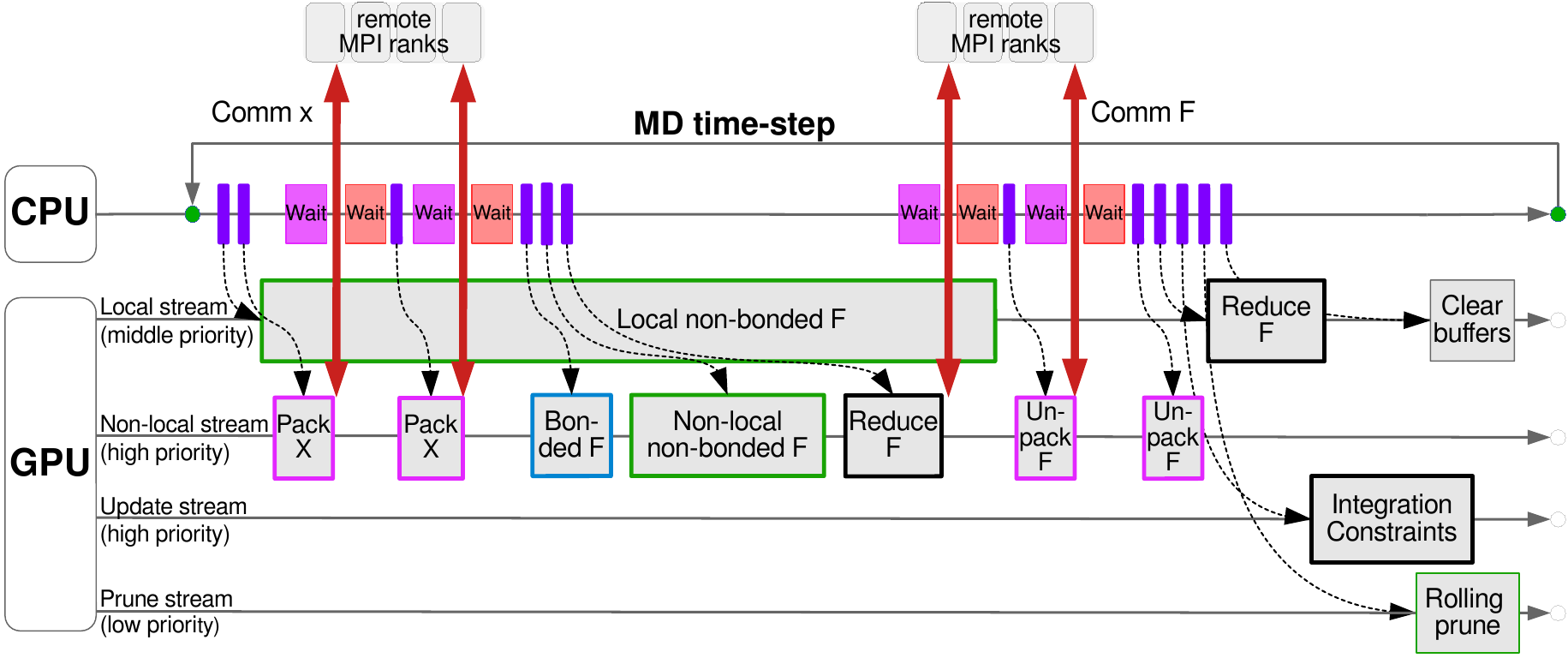}
    \caption{GROMACS GPU-resident MPI schedule illustrating a 2D DD setup with two X/F communication pulses. The top timeline represents the CPU MD time-stepping loop where purple boxes show kernel launches, magenta waits for GPU pack/unpack, while red boxes waits for MPI communication (red vertical arrows).
    The bottom timelines are GPU streams with a realistic representation of kernel execution showing partial overlap between local / non-local work and force communication / unpack exposed on the critical path.}
    \Description{A diagram showing CPU and GPU timelines of operations for the molecular dynamics workflow using MPI, where the CPU orchestrates communication and is forced into idle ``Wait'' states. This approach shows a clear separation between computation and communication, with the GPU performing distinct ``pack'' and ``unpack'' tasks, leading to performance bottlenecks.}
    \label{fig:mpi-schedule}
\end{figure}

GROMACS implements direct GPU communication support using GPU-aware MPI (shown in \figref{fig:mpi-schedule}) or alternatively based on thread-MPI which implements an event-driven, fully asynchronous communication setup. Since thread-MPI ranks are threads of the same process, GPU communication can employ direct DMA copies enqueued on GPU streams.
Unlike with regular MPI, this scheme can asynchronously launch both communication and computation for multiple iterations, overlapping GPU compute and launch. Dependencies are expressed as GPU events, and event synchronization is also used across multiple GPUs eliminating the need for CPU-based synchronization.

Together with a multi-GPU PME decomposition algorithm, GROMACS 2023 introduced cuFFTMp support \cite{gromacs2023} with a major leap in strong scaling, primarily due to the scalability of the NVSHMEM-based cuFFTMp.

\subsubsection*{Domain Decomposition Halo Exchange}

The GROMACS DD divides the simulation box into spatial regions (domains), with each MPI rank responsible for computing forces on atoms within its assigned domain. Non-bonded interactions are range-limited by a cutoff, hence atoms near domain boundaries require coordinates from neighboring domains to compute their forces. The halo exchange algorithm addresses this by communicating boundary atom data between adjacent domains \cite{Hess2008}.

The eighth-shell method \cite{Liem1991} employed by GROMACS uses forwarding-based (``staged'') communication: boundary data is forwarded through intermediate ranks rather than sent directly to all final consumers. We distinguish three terms to avoid ambiguity: (i) staged communication refers to this forwarding behavior, (ii) communication phases are the sequential z, then y, then x sweeps, and (iii) pulses are the per-dimension communication steps executed within a phase. GROMACS supports up to two pulses per dimension when second-neighbor communication is required and in cases where dynamic load balancing (DLB) produces a staggered DD grid \cite{Hess2008,gromacs_dd_manual}.
Currently, GROMACS does not support DLB in GPU-resident runs, hence staggered domain boundary case does not occur in this work. In addition, on heterogeneous machines, domain sizes are relatively large to provide sufficient parallelism per rank/GPU, hence the number of pulses per decomposition dimension is almost always one.

The halo communication proceeds in three sequential phases: first \textit{np}(z) pulses in the z-direction, then \textit{np}(y) in the y-direction, and finally \textit{np}(x) in the x-direction. 
This forwarding mechanism ensures that \textit{np}(x) + \textit{np}(y) + \textit{np}(z) communication steps suffice to exchange data with \textit{np}(x)\,$\times$\,\textit{np}(y)\,$\times$\,\textit{np}(z)\,$-$\,1 neighboring ranks while sending only directly required information \cite{Hess2008}. Although this method minimizes memory footprint and communication volume, it creates significant latency challenges in heterogeneous systems. Communication steps are coupled and each step requires expensive CPU-GPU synchronization, as GPU kernels must complete before MPI communication can proceed, and subsequent kernels cannot launch until data arrives. When scaling MD simulations, these sequential dependencies quickly become part of the critical path and ultimately limit strong scaling performance.

\begin{figure}
    \centering
    \includegraphics[width=\linewidth]{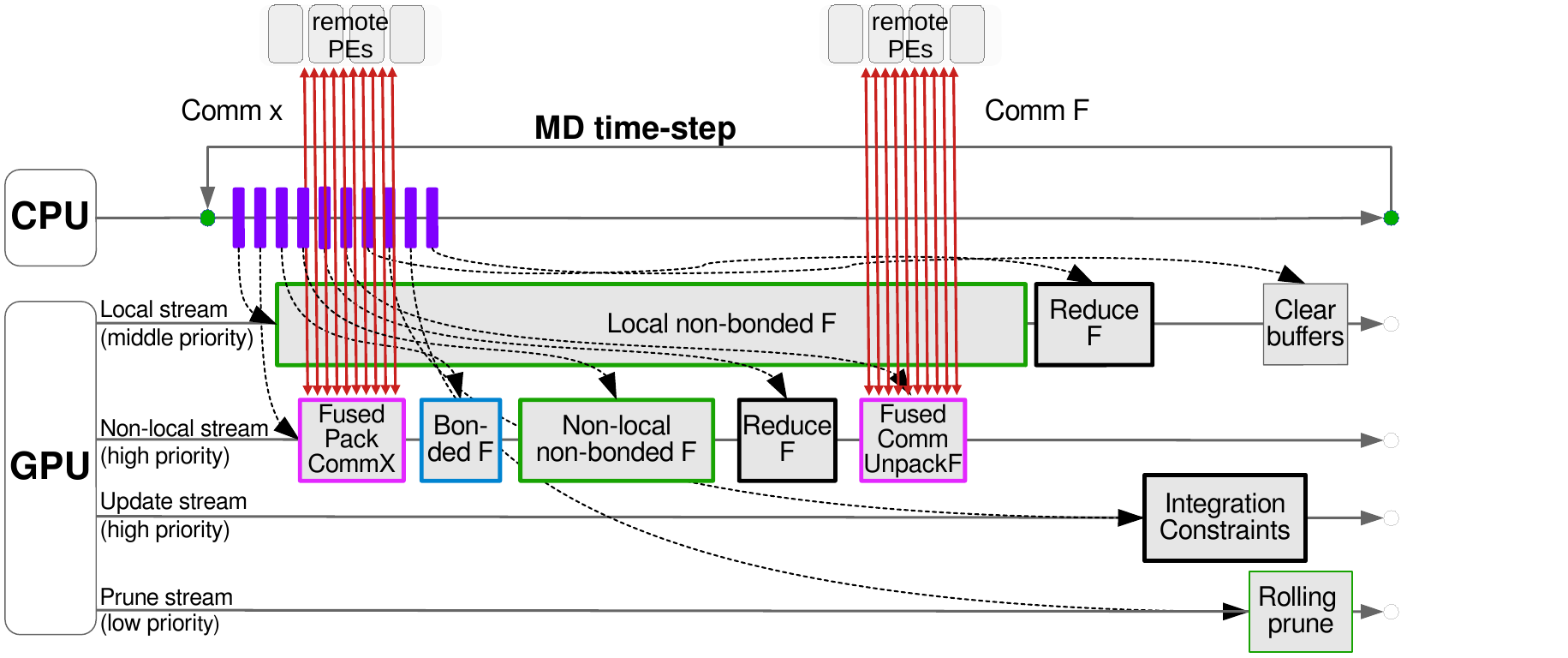}
    \caption{GROMACS GPU-resident NVSHMEM schedule illustrating a 2D DD setup (equivalent with \figref{fig:mpi-schedule}). The top timeline represents the CPU MD time-stepping loop where purple boxes show kernel launches which, unlike with MPI, can fully overlap with compute.
    The bottom timelines are GPU streams with a realistic representation of kernel execution showing complete overlap between local / non-local work, hence halo exchange is not on the critical path.}
    \Description{A diagram showing CPU and GPU timelines of operations for the molecular dynamics workflow using NVSHMEM, which eliminates the CPU's idle ``Wait'' states by enabling direct GPU-to-GPU data transfers. This allows for a superior overlap of computation and communication, further enhanced by streamlined ``fused'' operations on the GPU that combine packing and sending data.}
    \label{fig:nvhsmem-fused-schedule}
\end{figure}

\subsection{NVSHMEM}

NVSHMEM \cite{nvshmem_docs,potluri_gpu_2017} is a PGAS (Partitioned Global Address Space) programming API based on the OpenSHMEM standard \cite{openshmem_spec}, extending it to target NVIDIA GPU-based clusters. Inheriting from OpenSHMEM, NVSHMEM and similar GPU-optimized SHMEM implementations, 
like Intel SHMEM\cite{brooks_intel_2024} or rocSHMEM\cite{rocmrocshmem_2025}, create a global address space across multiple device memories, accessible with fine-grained operations directly from GPU kernels, thus enabling GPU-initiated communication.
Traditionally, SHMEM has been popular for applications with irregular communication patterns. More recently, particularly on GPUs, SHMEM-based communication has addressed latency challenges and enhanced the scalability of parallel algorithms. NVSHMEM allows offloading communication to the GPU, thereby consolidating the control path, reducing launch latencies, and enabling finer-grained overlap of communication and computation.
However, the requirement for a single global address space implies collective symmetric allocation across all PEs, which is at odds with rank specialization since GROMACS' PP and PME ranks have disjoint tasks and memory needs.

\section{GPU latency challenges in MD strong scaling}

The GPU architecture and programming models expose inherent latency. Reducing or avoiding the impact of these overheads has been a long-term goal of GROMACS \cite{Pall2020,alekseenko_gromacs_2024,nvidiablog2023}.
Key overheads include GPU kernel launches and memory copies (typically 2--10 µs), and to a lesser extent, event dependency management API calls (typically $<1$ µs).
Although often negligible in applications dominated by long-running compute kernels, fast-iterating applications are more sensitive to these overheads. Typical GROMACS time-steps require approximately 20 launch API calls and 30 event management calls. Their accumulated costs can be significant: at peak iteration rate it reaches over 50\% of GROMACS' CPU wall-time. This can throttle GPU execution, leaving idle gaps between kernels. 
By overlapping with GPU computation, such overheads can be mitigated as long as a sufficiently long sequence of asynchronous launches without CPU–GPU synchronization (provided total GPU work exceeds launch times). Further latency reductions are possible by using CUDA-graph scheduling for at least one entire time-step\cite{nvidiablog2023}.

With this in mind, in GROMACS CPU-GPU synchronization is kept to the minimum.
Long-term efforts have focused on optimization to maximize back-to-back time-step scheduling by reducing the frequency of “irregular” time-steps that require synchronization (e.g., I/O or CPU-based coupling algorithms).
In fully GPU-resident runs\footnote{Ones that do not have CPU computation contributing to the forces each time-step.}, typically tens to hundreds of time-steps are launched before CPU–GPU synchronization is required. 
Furthermore, completely avoiding per iteration CPU-GPU synchronization in a fully GPU-resident run has, until this work, only been possible on a single GPU or multi-GPU within a single node using thread-MPI. The same limitations apply to the GROMACS CUDA-graphs support.

Multi-node runs with MPI require multiple synchronizations per time-step, drastically reducing scope for overlap because the MPI standard lacks a seamless integration of communication into event dependency-driven GPU queue scheduling.
Consequently, the CPU needs to wait prior to each MPI call that operates on data produced by a GPU kernel. Similarly, compute kernels that consume MPI-communicated data can only be launched after ensuring the completion of the transfers. Examples are shown for a 2D halo exchange in \figref{fig:mpi-schedule}.
The GROMACS halo exchange requires multiple CPU-GPU synchronizations each time-step, often exposing resulting latencies on the critical path.
Furthermore, since MPI (stable implementations) cannot be invoked from GPU kernels, MD steps cannot be fully GPU-resident. With multiple halo exchange communication phases, pulses must be sequential, exacerbating overhead as simulation systems are decomposed along multiple dimensions.
These latency overheads, when not overlapped with computation and exposed on the critical path, limit strong scaling, which is why the GROMACS thread-MPI implementation can outperform regular MPI.
To improve multi-node scalability and further utilize GPU hardware's inherent latency-hiding capabilities, a communication API with better GPU integration is needed, and NVSHMEM is a good candidate.

\section{Related work}

Latency-optimizations using one-sided and PGAS communication techniques have been used for improving MD performance earlier on CPU-only systems. NAMD was codeveloped with the Charm++ framework 
for latency-tolerance \cite{Phillips2002}.
Previous work explored the benefits of SHMEM-based communication in GROMACS \cite{ReyTurHes2013} and LAMMPS \cite{tang_mpi_2015} reporting some performance benefits, but also drawbacks of symmetric allocation we encounter in our work. Straatsma et al.~avoids these with an ARMCI-based put-notify MD halo exchange implementation \cite{Straatsma2013}.

NVSHMEM has seen adoption in several communication backends and runtimes in recent years, for instance QUDA for lattice QCD \cite{Wagner_GTC_2020,gottlieb_two-link_2023}, CharminG a CHARM++-inspired GPU-resident runtime \cite{choi_charming_2021}, 
Livermore Big Artificial Neural Network runtime for spatial-parallel convolution \cite{VanEssen2015}, 
PETSc's PetscSF \cite{zhang_petsc_2021},
Kokkos Conjugate Gradient Solver, and
Kokkos Remote Space providing distributed memory support for Kokkos \cite{ciesko_kokkos_2023}. Several applications have also successfully used native NVSHMEM GPU-initiated communication including for GPU-based key-value stores \cite{chu_gpu_2021}, 
sparse triangular solver (SpTRSV) \cite{xie_fast_2020},
Graph Neural Networks (GNNs) \cite{wang_mgg_2023},
Jacobi 2D/3D and CG solvers \cite{ismayilov_multi-gpu_2023}.
While prior work in MD has used NVSHMEM to reduce memory usage at the cost of performance \cite{adjoua2021}, our work is the first to specifically leverage it to reduce latency and improve strong scaling.

\section{Implementation and optimizations}

\newcommand{\SINGLEWAIT}[3]{\STATE \textbf{#1 waits until} LD\_#2\_SYS(#3) = signalValue}
\newcommand{\BUFLENGTH}{2048}
\newcommand{\GLOBALTHREADIDX}{blockIdx.x $\times$ blockDim.x + threadIdx.x}

\begin{algorithm}
\footnotesize
\caption{Legend: Key data structures and sync objects}
\label{alg:legend}
\begin{algorithmic}[1]

\STATE \textbf{PulseData} (per pulse): \textit{sendRank}, \textit{recvRank}, \textit{sendSize}, \textit{recvSize}, \textit{atomOffset}, \textit{coordShift}, \textit{indexMap}, \textit{remoteCoordDst}, \textit{remoteForceDst}, \textit{remoteForceSrc}, \textit{sendBuf}, \textit{recvBuf}, \textit{blocksForPulse}
\STATE \textbf{pulseDataArray}: read-only array of \textbf{PulseData} entries in global pulse order [Z.., Y.., X..]
\STATE \textbf{CommContext}: \textit{totalPulses}, \textit{signal[p]}, \textit{sigVal}. Runtime path: NVLink when \textit{remoteCoordDst/remoteForceSrc} is non-null; otherwise InfiniBand with NVSHMEM put-with-signal. Per-pulse remote pointers (\textit{remoteCoordDst}, \textit{remoteForceSrc}) pre-populated before kernel launch using \textit{nvshmem\_ptr(ptr, remoteRank)} with the appropriate remote rank (\textit{sendRank} for puts, \textit{recvRank} for gets).
\STATE \textbf{Signals}: device-visible counters indexed by pulse id; sender \textbf{release-notifies} after completing all writes for that pulse; receiver \textbf{acquire-waits} before reading data that depends on that pulse
\STATE \textbf{Buffers}: \textit{coords}, \textit{forces}, and per-pulse staging buffers \textit{sendBuf}/\textit{recvBuf} when staging is required (InfiniBand path)
\STATE \textbf{Barriers}: \textit{indexMapLoadBarrier}, \textit{forceBufLoadBarrier} — intra-block barriers coordinating TMA-based shared-memory loads (NVLink path)
\STATE \textbf{Helper predicate}: isNVLinkAccess(ptr) $\coloneq$ (ptr $\neq$ null)

\end{algorithmic}
\end{algorithm}

Algorithm~\ref{alg:legend} summarizes the minimal types referenced by all pseudocode: \texttt{PulseData} carries per-pulse metadata, \texttt{CommContext} holds per-pulse signals and the selected transport path. \texttt{Buffers} are the main data (coordinates and forces, with optional staging buffers) that form the halos to be communicated, while \texttt{Signals} and \texttt{Barriers} are synchronization primitives. Per-pulse remote pointers (\texttt{remoteCoordDst}, \texttt{remoteForceSrc}) are assumed pre-populated with \texttt{nvshmem\_ptr()} before kernel launch. Here, names ending with \textit{Dst} denote the peer-side destination for puts (sendRank), and names ending with \textit{Src} denote the peer-side source for gets (recvRank); all \textit{remote*} symbols are device pointers to peer GPU memory.

\begin{algorithm}
\footnotesize
\caption{GPU-resident time-step skeleton (fused NVSHMEM halo exchange)}
\label{alg:timestep-skeleton}
\begin{algorithmic}[1]
\STATE launch \textit{Local Non-bonded F} on local stream
\STATE launch \textit{FusedPackCommX} on non-local stream  \COMMENT{Coordinate halo: pack, NVLink TMA store or InfiniBand put-to-remote, then notify}
\STATE launch \textit{Bonded F} on non-local stream
\STATE launch \textit{Non-Local Non-bonded F} on non-local stream
\STATE launch \textit{FusedCommUnpackF} on non-local stream \COMMENT{Forces halo: NVLink use get-from-remote or InfiniBand use put-to-remote; notify; then unpack}
\STATE launch \textit{Integration, constraints} on update stream
\STATE launch other per-step work: \textit{ReduceF}, \textit{Rolling prune}, \textit{Clear buffers}
\STATE proceed to next step without CPU-GPU sync (GPU-resident schedule)
\end{algorithmic}
\end{algorithm}

\begin{algorithm}
\footnotesize
\caption{Kernel: FusedPackCommX}
\label{alg:fused_pack_nvshmem_part1}
\begin{algorithmic}[1]

\REQUIRE coords, pulses: PulseData[\,], ctx: CommContext
\STATE blocks with the same blockIdx.y cooperate on the current pulse p (this loop iteration); blockIdx.x partitions chunks within that pulse
\FOR{p $\gets$ blockIdx.y \TO ctx.totalPulses-1 \textbf{step} gridDim.y}
    \STATE meta $\gets$ pulses[p];
    \STATE outBuf $\gets$ (isNVLinkAccess(meta.remoteCoordDst)) ? shared-mem scratch : meta.sendBuf; needLastBlock $\gets$ (meta.blocksForPulse $>$ 1)
    \STATE \textbf{packWithDeps}(meta, ctx, p, outBuf) 

    \IF{isNVLinkAccess(meta.remoteCoordDst)} 
        \STATE as chunks are ready: \textbf{tma\_st}(outBuf $\rightarrow$ meta.remoteCoordDst + meta.atomOffset) \COMMENT{async; issued by thread 0 of each warp (warp leader)}
   \ENDIF
   \STATE syncAndCommWithDeps(\textsc{DATA}, needLastBlock, ctx, meta, p, pulses, outBuf, null, coords) \COMMENT{NVLink vs InfiniBand handled inside} 
    
\ENDFOR
\end{algorithmic}
\end{algorithm}

\begin{algorithm}
\footnotesize
    \caption{Device Function: packWithDeps}
    \label{alg:dependency_aware_pack_loop}
    \begin{algorithmic}[1]
    \REQUIRE meta: PulseData, ctx: CommContext, currentPulse p, outBuf
    \STATE pack all entries with meta.indexMap[i] $<$ meta.depOffset into outBuf (apply meta.coordShift as needed)
    \STATE elect a leader at the chosen scope (warp leader for NVLink path, or threadIdx.x=0 otherwise)
    \STATE leader: for k $\gets$ p-1 \textbf{down to} firstDependentPulse: \textbf{acquire\_wait}(ctx.signal[k] == ctx.sigVal)  
    \STATE scope\_barrier() \COMMENT{SYNCWARP() for warp scope(NVLINK); SYNCTHREADS() for block scope (InfiniBand)}
    \STATE pack remaining entries with meta.indexMap[i] $\ge$ meta.depOffset into outBuf
\end{algorithmic}
\end{algorithm}

\subsection{Coordinate halo exchange}
\subsubsection*{Baseline (serialized pulses)} A straightforward de\-pen\-den\-cy-{}pre\-serv\-ing implementation processes pulses sequentially: for each dimension (z, then y, then x), the implementation packs and communicates the data for one pulse before proceeding to the next. This is how the MPI formulation ensures forwarding dependencies, but it serializes communication and relies on distinct pack/unpack kernels per pulse (Fig.~\ref{fig:mpi-schedule}).

\subsubsection*{Fused (parallel pulses)} 
%
In contrast, our fused design launches a single kernel that processes all pulses in parallel to maximize concurrency and reduce kernel-launch overhead (Algorithm~\ref{alg:fused_pack_nvshmem_part1}). The kernel assigns a separate threadblock to each pulse and manages only the necessary forward-direction dependencies using fine-grained, per-pulse signaling and scope barriers across \texttt{packWithDeps} (Algorithm~\ref{alg:dependency_aware_pack_loop}) and \texttt{syncAndCommWithDeps} (Algorithm~\ref{alg:sync_and_comm_with_deps}). 

%
 The core of this design is dependency partitioning (Algorithm~\ref{alg:dependency_aware_pack_loop}). We split the index map at the offset \texttt{depOffset}; entries below this are packed and enqueued for transfer immediately, while processing of dependent entries waits for the previous pulse's signal. This strategy ensures that independent coordinate packing and communication proceeds without delay, and dependent work to only wait as long as necessary, allowing all pulses to advance in parallel while respecting the required ordering.
 
 To determine \textit{firstDependentPulse}, we define a global pulse order by concatenating dimensions in Z\,$\rightarrow$\,Y\,$\rightarrow$\,X order, omitting dimensions not present in the current DD, e.g.~for one pulse per dimension ordered [z0, y0, x0]: \textit{firstDependentPulse}(z0)=none; \textit{firstDependentPulse}(y0)=z0; \textit{firstDependentPulse}(x0)=y0.

%
 The kernel dynamically adapts its communication strategy at runtime based on the interconnect. NVLink accessibility is established early, allowing zero-copy direct remote memory access. We use a fine-grained, sender-driven approach with direct remote memory writes via the Tensor Memory Accelerator (TMA)\cite{hopper_tuning_guide}. For sending across InfiniBand, it uses coarsened put-based communication with per-pulse staging buffer to amortize latency and signaling overhead. Signals use the concise ctx.signal and expected value ctx.sigVal.
%



%
On NVLink, we pipeline packing with the warp-level TMA remote stores while preserving pulse dependencies (Algorithm~\ref{alg:dependency_aware_pack_loop}). Each warp within a threadblock packs a dedicated chunk of coordinates into shared memory, respecting alignment requirements for TMA coalescing of 128-bytes and float3 layout. As a chunk is packed, an elected warp leader issues the asynchronous TMA store operation, while other warps can continue packing. This approach decouples progress between warps, avoiding block-wide synchronization, minimizing latency to TMA issue, and effectively pipelining packing and remote stores. We use \texttt{cuda::ptx::cp\_async\_bulk} to drive the TMA engine~\cite{cccl_cp_async_bulk}. TMA autonomously handles transfer, freeing SM resources for computation, maximizing GPU utilization.

%
For InfiniBand, the kernel still uses \texttt{packWithDeps} (Algorithm~\ref{alg:dependency_aware_pack_loop}) to partition work based on dependencies. However, it packs coordinates into a staging buffer in global memory before issuing remote put via nvshmem\_float\_put\_signal\_nbi. As before, independent data is packed immediately, while dependent data from previous pulses is waited on through signal-based synchronization, overlapping the packing of independent data with dependency-resolution.




\begin{algorithm}
\footnotesize
\caption{Device Function: syncAndCommWithDeps}
\label{alg:sync_and_comm_with_deps}
\begin{algorithmic}[1]

\REQUIRE mode $\in$ \{\textsc{DATA}, \textsc{DEP\_MGMT}\}, needLastBlock, ctx: CommContext, meta: PulseData, p, pulses[\,], outBuf, forces, coords
\STATE block\_barrier(); only \textbf{threadIdx.x = 0} proceeds to notify
\IF{needLastBlock}
    \STATE old $\gets$ atomicIncReleaseGpu(blockCompletionCounter[p]) \COMMENT{GPU-scope release increment}
    \STATE \textbf{if} old $\neq$ meta.blocksForPulse-1 \textbf{return}
\ENDIF
\STATE isDepMgmt $\gets$ (mode == \textsc{DEP\_MGMT}); prev $\gets$ pulses[p-1]; curr $\gets$ meta
\STATE receiverSignal $\gets$ isDepMgmt ? ctx.signal[p-1] : ctx.signal[p]; sendRank $\gets$ isDepMgmt ? prev.sendRank : curr.sendRank
\IF{isDepMgmt}
    \STATE for k $\gets$ p+1 \textbf{to} ctx.totalPulses-1: \textbf{acquire\_wait}(ctx.signal[k] == ctx.sigVal) \COMMENT{ensure all subsequent pulses completed before moving forward}
\ENDIF
\IF{coords} 
    \STATE src $\gets$ outBuf; dest $\gets$ coords + curr.atomOffset; size $\gets$ curr.sendSize; isNvLink $\gets$ isNVLinkAccess(curr.remoteCoordDst); hasDataWrites $\gets$ isNvLink; useNvshmemPut $\gets$ (!isNvLink)
\ELSE 
    \IF{isDepMgmt} 
        \STATE src $\gets$ forces + prev.atomOffset; dest $\gets$ prev.recvBuf; size $\gets$ prev.sendSize
        \STATE isNvLink $\gets$ isNVLinkAccess(prev.remoteForceSrc); hasDataWrites $\gets$ isNvLink; useNvshmemPut $\gets$ !isNVLinkAccess(prev.remoteForceDst)
    \ELSE 
        \STATE src $\gets$ forces + curr.atomOffset; dest $\gets$ outBuf; size $\gets$ curr.sendSize
        \STATE isNvLink $\gets$ isNVLinkAccess(curr.remoteForceSrc); hasDataWrites $\gets$ false; useNvshmemPut $\gets$ !isNVLinkAccess(prev.remoteForceDst)
    \ENDIF
\ENDIF
\STATE \textbf{if} isNvLink \textbf{then} (hasDataWrites ? \textbf{system\_release\_store}(receiverSignal, ctx.sigVal) : \textbf{system\_relaxed\_store}(receiverSignal, ctx.sigVal))
\STATE \textbf{if} useNvshmemPut \textbf{then} nvshmem\_float\_put\_signal\_nbi(dest, src, size, receiverSignal, ctx.sigVal, sendRank) \COMMENT{issued by last-arriving block in general; global threadIdx=0 on initial forces send}

\end{algorithmic}
\end{algorithm}

For optimal performance, kernel resource utilization is dynamically adjusted based on workload and communication topology.
For TMA-enabled NVLink transfers it uses \texttt{(sendSize / bufLength)} threadblocks to match TMA transfer granularity. For the non-NVLink case (including InfiniBand), it uses the full grid dimension with a grid-strided loop to maximize parallel processing. This maintains data consistency through careful ordering of packing, signaling, and synchronization operations, achieving optimal communication across diverse hardware configurations. Notification and communication are issued by \texttt{syncAndCommWithDeps} (Algorithm~\ref{alg:sync_and_comm_with_deps}), which enforces per-pulse ordering and performs the transport-specific operation (NVLink vs InfiniBand; see §\ref{sec:fused-signaling}).

\subsection{Force halo exchange}
\begin{algorithm}
\footnotesize
\caption{Kernel: FusedCommUnpackF}
\label{alg:fused_force_unpack}
\begin{algorithmic}[1]

\REQUIRE forces[], pulses: PulseData[\,], ctx: CommContext, accumulate
\FOR{p $\gets$ ctx.totalPulses - 1 - blockIdx.y  \textbf{down to} 0  \textbf{step} -gridDim.y}
    \STATE meta $\gets$ pulses[p] \COMMENT{reverse send/recv roles vs coordinates}
    \IF{ p = ctx.totalPulses-1 \AND (\GLOBALTHREADIDX) = 0 }
        \STATE syncAndCommWithDeps(\textsc{DATA}, false, ctx, meta, p, pulses, meta.recvBuf, \&forces[meta.atomOffset], null)
    \ENDIF
     \STATE dataSize $\gets$ (meta.remoteForceSrc ? meta.recvSize : meta.sendSize); bufLength $\gets$ \BUFLENGTH; chunkOffset $\gets$ blockIdx.x $\times$ bufLength; chunkSize $\gets$ min(bufLength, dataSize - chunkOffset)
    \STATE Initialize shared memory buffers and barriers (\textit{indexMapLoadBarrier}, \textit{forceBufLoadBarrier})
    \IF{threadIdx.x $==$ 0 \AND chunkOffset $<$ dataSize}
        \STATE \textbf{tma\_ld}(smemMap $\leftarrow$ meta.indexMap + chunkOffset, chunkSize, indexMapLoadBarrier)
        \STATE \textbf{acquire\_wait}(ctx.signal[p] == ctx.sigVal)
    \ENDIF
    \IF{isNVLinkAccess(meta.remoteForceSrc)}
        \STATE \textbf{tma\_ld}(smemF $\leftarrow$ meta.remoteForceSrc + meta.atomOffset + chunkOffset, bufLength, forceBufLoadBarrier) \COMMENT{issued by threadIdx.x = 0}
        \STATE forceBufLoadBarrier.arrive\_and\_wait()
    \ENDIF
    \STATE indexMapLoadBarrier.arrive\_and\_wait()
    \STATE in parallel across threads: for each received entry (from shared-memory chunk on NVLink or staged buffer \textit{meta.recvBuf} on InfiniBand), map to target atom index using the index map and write/accumulate into \textit{forces} (use atomicAdd when accumulating)
    \IF{p $>$ 0}
        \STATE syncAndCommWithDeps(\textsc{DEP\_MGMT}, true, ctx, meta, p, pulses, null, \&forces[0], null)
    \ENDIF
\ENDFOR

\end{algorithmic}
\end{algorithm}

This implementation (Algorithm~\ref{alg:fused_force_unpack}) fuses halo‑force communication and unpacking by enabling parallel unpacking across all pulses using \texttt{atomicAdd}, thereby minimizing wait times for data that depend on prior pulses. It begins with the last pulse's computed forces, available after the non-local non-bonded force kernel completes, and works backwards through the dependency chain. It uses separate threadblocks for each pulse, with communication following the bidirectional halo exchange pattern where ranks that provided coordinates now receive the computed forces for those atoms. For NVLink, we uses device-initiated gets into shared memory (\textit{tma\_ld}). For InfiniBand, we use staged buffers with signal waits. After transport-specific setup, we use a single parallel accumulation step common to both paths, mapping each received entry through the index map and writing or accumulating into \textit{forces}.

For NVLink, the kernel leverages the TMA engine to perform asynchronous bulk gets from remote GPU global memory into local shared memory. 
The index map is loaded first, followed by an acquire-wait on the peer's signal that it is safe to load forces, followed by a bulk load of the force chunk; both loads are issued by a single leader and synchronized with shared-memory barriers.
We use a get approach because it allows the receiver to drive the schedule and consume data when its threads are ready. This avoids remote atomics or pre-accumulation on the sender side, keeps ownership of accumulation local, and naturally fits the TMA global-to-shared load path that feeds a shared-memory working set before accumulating into the destination \textit{forces} array.

For InfiniBand, the kernel waits on the peer signal and unpacks from the staged buffer \texttt{recvBuf}. Upon arrival of forces it unpacks using \texttt{atomicAdd}, accumulating in parallel without having to wait for prior pulses. Only the last-completing threadblock per pulse waits for dependencies before forwarding or signaling availability to subsequent pulses. Per-pulse device-side signals gate consumption, and shared-memory barriers (\textit{indexMapLoadBarrier}, \textit{forceBufLoadBarrier}) coordinate TMA-based loads before accumulation.

The force halo exchange uses \texttt{syncAndCommWithDeps} (Algorithm~\ref{alg:sync_and_comm_with_deps}) for dependency forwarding
in dependency-management (DEP\_MGMT) mode (details in §\ref{sec:fused-signaling}). It handles the dependency chain where each pulse must wait for prior pulses to complete before forwarding accumulated forces with fused receiver notification. This communication design maximizes performance by making newly computed forces immediately available while other pulses proceed in parallel, shortening the critical path. 



\subsubsection*{Fused signaling and memory ordering} \label{sec:fused-signaling}

An important innovation in our design, unlike typical NVSHMEM implementations, is to combine data packing/transfer with receiver notification in a single kernel via \texttt{syncAndCommWithDeps} (Algorithm~\ref{alg:sync_and_comm_with_deps}). Each participating threadblock performs a GPU-scope release atomic increment on completion; the last completing block issues the communication and the receiver notification operation. Over InfiniBand we use \texttt{nvshmem\_float\_put\_signal\_nbi} (non-blocking put-with-signal). Over NVLink, notification uses system-scoped release stores; \texttt{system\_release\_store} ensures visibility of preceding data writes, while \texttt{system\_relaxed\_store} is used when no prior writes need flushing (e.g., the first pulse of the force send), these are our thin wrappers around PTX \footnote{st.release.sys.global  and st.relaxed.sys.global}. Emitting system-scope operations only from the last block minimizes overhead while preserving ordering. We rely on the C++11 acquire-release model and CUDA PTX memory semantics for correctness across device and system scope~\cite{libcudacxx_memory_model,ptx_isa_docs}.

\subsection{Integration into the GPU-resident schedule}

Our NVSHMEM halo exchange preserves GROMACS' event-driven GPU execution but eliminates CPU–GPU synchronization bottlenecks inherent to the CPU-initiated MPI-based approach as illustrated in Figs.~\ref{fig:mpi-schedule} and \ref{fig:nvhsmem-fused-schedule}. The overall GPU-resident control flow is shown in Algorithm~\ref{alg:timestep-skeleton}. This fused design enables a fully GPU-resident halo exchange where all kernels 
can be launched early in the time-step without intermediate synchronization.
This replicates thread-MPI's ability to overlap kernel launch with compute, and extends its unified device-driven control path (fewer CPU–GPU synchronizations) benefits to multi-node configurations, while also eliminating its GPU copy engine launch overheads. 
The NVSHMEM halo exchange kernels use the same high-priority streams as the MPI implementation leaving kernel scheduling behavior unchanged, but the kernel launch overhead is reduced by issuing one launch per coordinate/force exchange rather than one per pulse.

NVSHMEM symmetric allocations are handled transparently during initialization, and buffer sizes adjust to DD requirements, though thanks to the GROMACS' over-allocation strategy, resizing is rarely required.
However, a limitation currently prevents combining our halo exchange with rank specialization when multiple PME ranks use cuFFTMp. 
The issue arises because NVSHMEM's COMM\_WORLD-wide symmetric allocation model prevents selective PP/PME participation: PP-only symmetric destination buffers would require redundant PME allocations and vice versa. With cuFFTMp 
such allocations are not user-controllable. NVSHMEM's symmetric allocation requirement applies to the destination buffer; the source buffer can be non-symmetric allocation registered using nvshmemx\_buffer\_register \cite{nvshmemx_buffer_register_docs}. The implementation maintains compatibility with CUDA graph capture, enabling time-steps including NVSHMEM communication to be captured for latency reduction. 

\subsection{Other critical path optimizations}
During performance analysis, we identified that the original GROMACS heterogeneous schedule was suboptimal, delaying the critical path.
The dynamic pair list pruning algorithm was designed before the GPU-resident mode was added, and pruning was scheduled after the force kernels to overlap with CPU-side integration \cite{Pall2020}.
However, in the GPU-resident mode, this scheduling can cause the prune kernel
to execute early, blocking the integration and consequently delaying the critical path of the following step. To resolve this, we revised the end of time-step schedule in two ways. 
First, we moved the prune kernels to a dedicated, low-priority GPU stream, relaxing its prior implicit stream-ordered dependencies and shifting their launch to the end of the step. Second, we introduced a third medium priority stream for the local reduction and update tasks, ensuring that they can progress by preempting pruning (Figs.~\ref{fig:mpi-schedule}, \ref{fig:nvhsmem-fused-schedule}).
Collectively, these optimizations improve performance by up to 10\% for both MPI and NVSHMEM implementations, with slightly greater benefits typically observed for the latter.

\subsection{Proxy thread affinity challenges}

While GROMACS uses per-thread affinities to optimize data locality, this creates a challenge with NVSHMEM. 
To avoid resource contention, it is essential not only that a free core is available for the NVSHMEM proxy thread, but also that the proxy thread is not pinned to an already busy core. 
Failing to ensure this can cause significant performance degradation, we observed up to 50x slowdown in our multi-node tests.

GROMACS threads in a rank are generally pinned to a contiguous set of cores on the system, usually a single NUMA region or socket. 
Existing thread-pinning mechanisms, however, are insufficient. 
GROMACS's internal \verb|-pin| option cannot underpopulate ranks to leave cores free for the NVSHMEM proxy thread (or other runtime threads).
External pinning via OpenMP is a possible alternative using \verb|OMP_PLACES=cores| \verb|OMP_PROC_BIND=close| and \verb|OMP_NUM_THREADS| fewer than the number of cores per rank, but its behavior is unreliable and inconsistent.
The proxy thread inherits its affinity from the main thread at the time of calling \verb|nvshmem_init|, and the timing of this inheritance varies by runtime.  From our experiments, the GNU OpenMP runtime sets affinities very early (at libgomp library load), pinning the main and proxy threads to the same core. In contrast, LLVM OpenMP sets them late enough so that the proxy thread correctly remains unpinned, achieving the desired behavior.

To create a reliable contention-free solution,
 we modified GROMACS to use \verb|OMP_NUM_THREADS-1| threads in its parallel regions. The last, unused thread is used to initializing NVSHMEM, thus pinning the proxy thread to a free core.\footnote{This requires NVSHMEM to be initialized in NVSHMEM\_THREAD\_SERIALIZED mode and that the thread calling \texttt{nvshmem\_init} accesses the same CUDA context as the main thread.} In our benchmarks showed no benefit from this thread level pinning over only per-rank pinning. This is likely due to low OS noise and the fact that any migrated threads are not crossing socket boundaries. 

\section{Performance evaluation and results} 

\subsection{Benchmark Systems and Inputs}

Our performance evaluation was conducted on the NVIDIA Eos supercomputer\cite{nvidiablog2024}. Eos consists of 576 DGX H100 nodes, each equipped with two Intel Xeon 8480C CPUs and eight H100s, for a total of 4,608 GPUs. Intra-node it uses NVLink 4.0 and NVSwitch interconnect topology to connect GPUs\cite{nvidia_fabric_manager_h100}.  
The system employs NVIDIA ConnectX-7 NDR 400G InfiniBand networking, 
in a rail optimized, non-blocking full fat-tree topology. Running Ubuntu 22.04 with CUDA Toolkit 12.6.3, CUDA Driver 535.129.03, GCC 13.1.0, Open MPI 4.1.7rc1, UCX 1.19.0, and NVSHMEM 3.1.7 (with IBRC transport).

To evaluate performance in different scaling regimes, we use the ``grappa'' benchmark set, which consists of water ethanol mixture, and sizes between $45000$ and $46$ million atoms. The computational workload in these benchmark systems closely resembles that of the typical biomolecular simulation, with the slight difference that these systems are more homogeneous. We use a reaction-field model for electrostatics to allow focusing the analysis on short-range interactions and halo exchange, for detail see \cite{doijade_2025}.

\begin{figure}
    \centering
    \includegraphics[width=\linewidth]{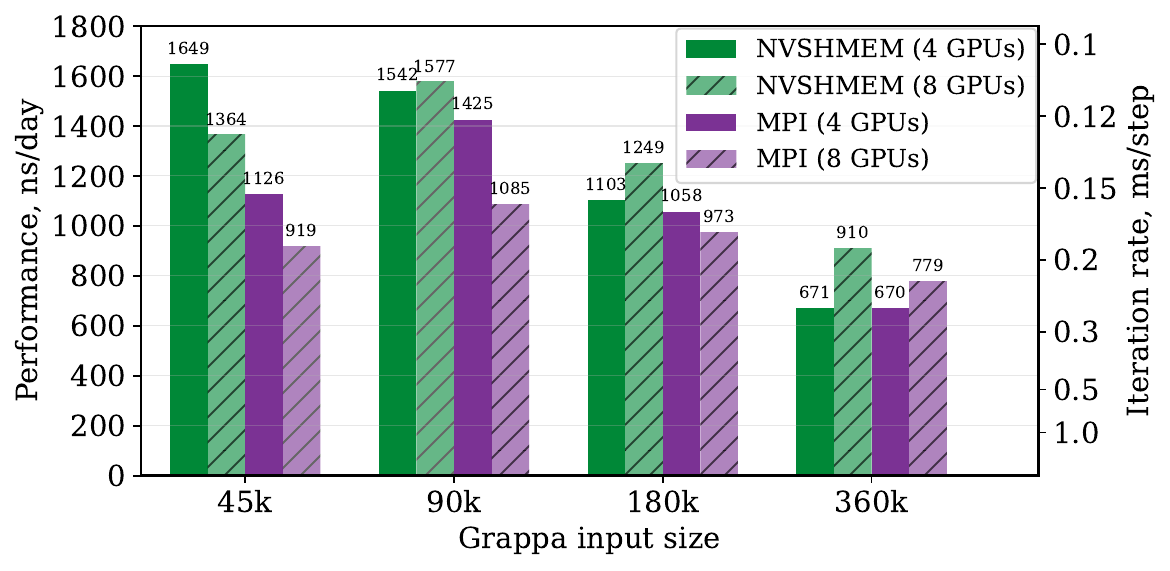}
    \caption{Intra-node MPI and NVSHMEM comparison on 4/8 GPUs showing simulation performance (ns/day) and iteration rate (ms/step). The iteration rate is the average wall-time per time-step, computed across all time-steps.
    }
    \Description{A clustered bar chart comparing the performance of NVSHMEM (green bars) and MPI (purple bars) using 4 GPUs (solid bars) and 8 GPUs (hatched bars). The chart shows ``Performance`` in nanoseconds per day on the primary Y-axis versus four different Grappa input sizes on the X-axis. In all tested conditions, NVSHMEM shows higher performance than MPI. A general trend shows that performance decreases for all methods as the input size increases. For both NVSHMEM and MPI, the 4-GPU configuration outperforms the 8-GPU configuration for Grappa 45k system. For Grappa 90k and Grappa 180k, NVSHMEM performance improves when going from 4 GPUs to 8 GPUs, while MPI performance decreases. For Grappa 360k system, both MPI and NVSHMEM perform better on 8 GPUs than on 4 GPUs. For each tested configuration, NVSHMEM has better performance than MPI on the same input and hardware.}
    \label{fig:intra-node-scaling-comparison}
\end{figure}

\subsection{Strong scaling} 

Strong scaling performance shows distinct benefits of our NVSHMEM based approach over MPI across the evaluated system sizes, with performance characteristics varying by system size and node count.

For evaluating intra-node scaling we use up to 8 GPUs in DGX-H100. NVSHMEM shows significant advantages, particularly for smaller systems (\figref{fig:intra-node-scaling-comparison}). The 45k system demonstrates the largest NVSHMEM performance gains with 46\% at 4 GPUs (1649 vs 1126 ns/day). Increasing system size, NVSHMEM maintains a moderate advantage, at 180k 4\% on 4 GPUs (1103 vs 1058 ns/day) and 28\% on 8 GPUs (1249 vs 973 ns/day). At 360k  performance converges on 4 GPUs (671 vs 670 ns/day) but NVSHMEM maintains 17\% advantage on 8 GPUs (910 vs 779 ns/day). This trend reflects the transition from communication-bound to compute-bound regimes when scaling intra-node, where latency optimizations using TMA bulk transfers provide the greatest benefit for smaller simulation systems with insufficient compute to hide communication overhead.

\begin{figure}
    \centering
    \includegraphics[width=\linewidth]{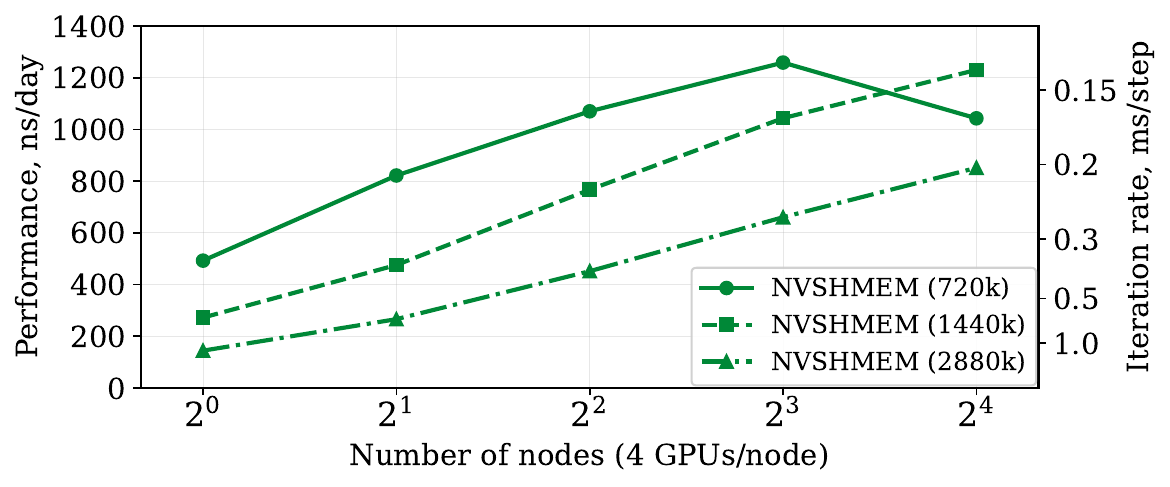}
    \caption{NVSHMEM strong scaling on 
    a GB200 NVL72 
    showing simulation performance (ns/day) and iteration rate (ms/step) as a function of the number of nodes. The iteration rate is the average wall-time per time-step, computed across all time-steps.
    }
    \Description{A line graph showing performance scaling of NVSHMEM with an increasing number of nodes for three system sizes. The horizontal x-axis represents the number of nodes, from 1 node to 16 nodes. The primary vertical y-axis on the left shows ``Performance'' in nanoseconds per day, while the secondary y-axis on the right shows ``Iteration rate'' in milliseconds per step on an inverted scale. The graph presents three data series corresponding to different system sizes: NVSHMEM (720k): A solid line with circle markers. NVSHMEM (1440k): A dashed line with square markers. NVSHMEM (2880k): A dash-dot line with triangle markers. Generally, performance increases for all three systems as the number of nodes increases. The performance for the 720k system peaks at 8 nodes (approximately 1250 ns/day) and then decreases at 16 nodes. The other two systems show a continuous increase in performance across all node counts, with smaller systems having higher performance.}
    \label{fig:mnnvl-scaling-comparison}
\end{figure}

We also evaluated scaling performance of our NVLink-optimized implementation on an NVIDIA GB200 NVL72 \cite{gb200_nvl72_datasheet} multi-node NVLink (MNNVL) cluster in a 36x2 configuration\cite{nvidia_gb200_nvl_tuning_guide} (hardware and software setup detailed in the Artifact Description $A_1$) 
for the 720k, 1440k, and 2880k grappa systems (\figref{fig:mnnvl-scaling-comparison}).
Using single-node performance as baseline (492 ns/day for 720k, 272 ns/day for 1440k), the 720k system demonstrates strong parallel efficiency with NVSHMEM, 84\% on 2 nodes (8 GPUs), 55\% on 4 nodes (16 GPUs), and 32\% on 8 nodes (32 GPUs). The 1440k system shows even better scaling with 88\% efficiency on 2 nodes, 71\% on 4 nodes, and 48\%  on 8 nodes, showcasing the benefits of our fine-grained zero-copy NVlink-specific algorithm effectiveness on also larger number of GPUs across a multi-node NVLink network.
Although we are unable to show MPI scaling here\footnote{Due to time constraints, and software setup challenges, MPI could not be tested reliably on this system.}, a comparison based on early data suggests up to 2x higher performance with NVSHMEM at scale.

\begin{figure}
    \centering
    \includegraphics[width=\linewidth]{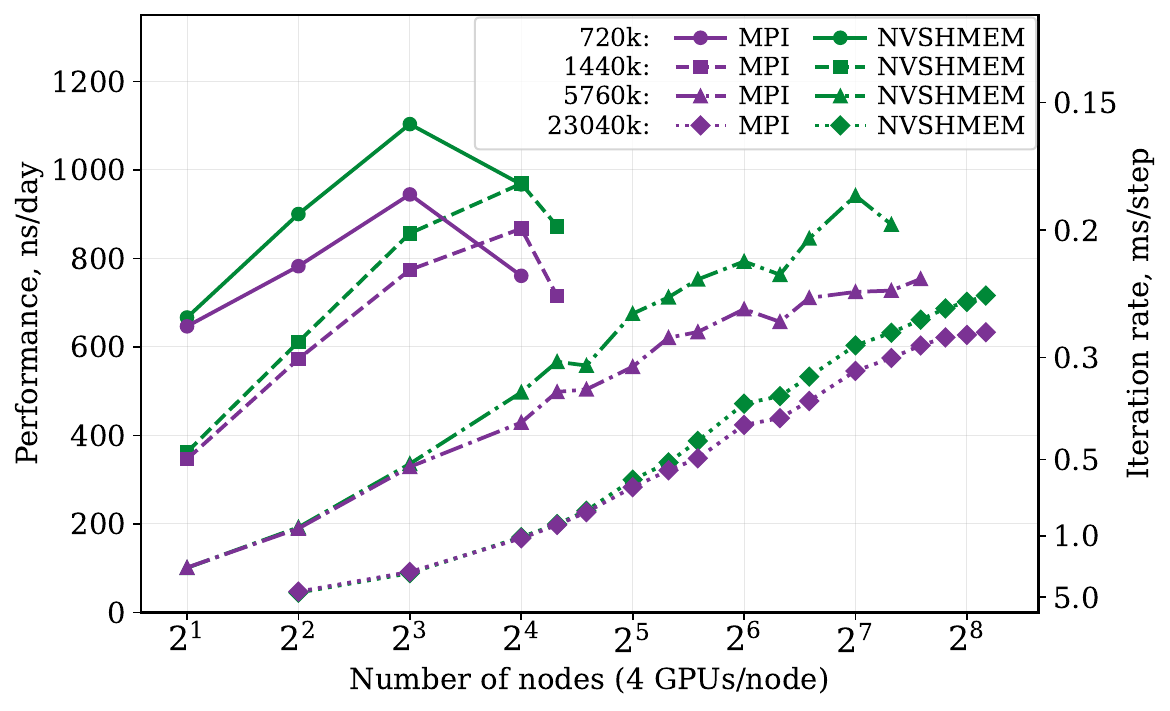}
    \caption{Multi-node MPI and NVSHMEM strong scaling showing simulation performance (ns/day) and iteration rate (ms/step) as a function of the number of nodes. The iteration rate is the average wall-time per time-step, computed across all time-steps. 
    }
    \Description{Line graph comparing strong scaling performance of MPI and NVSHMEM implementations across multiple node counts. The x-axis shows the number of nodes, each with 4 GPUs, ranging from 2 to 256. The left Y-axis shows simulation ``Performance'' in nanoseconds per day, from 0 to 1300. The right Y-axis shows ``Iteration rate'' in milliseconds per step. Four problem sizes are included: 720k; 1440k; 5760k; and 23040k particles. For each size, results are plotted for both MPI and NVSHMEM. Trends show that performance in nanoseconds per day generally increases with node count, while iteration rate per step decreases. NVSHMEM curves consistently achieve higher performance (more ns/day and lower ms/step) than MPI at each problem size. The relative scaling advantage of NVSHMEM is most pronounced at larger node counts and larger problem sizes.}
    \label{fig:strong-scaling-comparison}
\end{figure}

For multi-node evaluation on Eos, we use four of the eight H100 GPUs per node as this configuration more closely resembles typical HPC system deployments, providing 8-1152 total GPUs across the tested configurations. NVSHMEM consistently outperforms MPI for smaller systems (\figref{fig:strong-scaling-comparison}): for 720k  17\% higher performance (1103 vs 944 ns/day on 8 nodes) with superior parallel efficiency (42\% vs 37\%). For 1440k this advantage is maintained with 12\% higher peak performance and 33\% vs 31\% efficiency on 16 nodes. The 5760k system shows a continued parallel scaling efficiency advantage (15\% vs 11\% at 128 nodes) with NVSHMEM achieving 1.3x higher performance than MPI on 128 nodes.



For larger systems at low node counts, MPI marginally outperforms NVSHMEM, where NVSHMEM's SM resource-sharing overhead is more visible on compute‑intensive workloads. For multi-node configurations, our scaling evaluation spans a wide range from ~10k to 1440k atoms per GPU over many node counts. The fundamental scaling limit is determined by the available parallelism per GPU: performance at scale becomes primarily bound by GPU under‑utilization; with fewer than 10–25k atoms per GPU there is not enough compute work to saturate modern high‑SM‑count ($>$100 SMs) H100s. Approaching this observed scaling limit, GPU under‑utilization becomes the dominant factor in the loss of parallel efficiency and cannot be compensated for by latency‑optimized communications alone. This is supported by the observation that strong scaling peaks near 20k atoms per GPU both across NVLink (e.g.,~the 90k system on 4--8 GPUs intra‑node) and NVLink + InfiniBand inter-node (e.g., the 23040k system at 288 nodes). Consistent with this, at larger sizes and higher node counts we see continued efficiency advantages and higher absolute performance for NVSHMEM at scale (e.g.,~23040k: 716 vs 633~ns/day at 288 nodes), while MPI can retain a slight advantage at lower node counts (Fig.~\ref{fig:strong-scaling-comparison}).

These results demonstrate the superior communication efficiency of our NVSHMEM-based approach, which shows significant advantages in most scaling regimes, particularly at scale. In compute-dominated scenarios (large inputs on few nodes), MPI holds a minor 1-3\% performance advantage. This highlights that in our case the GPU SM resource demands of NVSHMEM's device-initiated communication impose only a minimal overhead. It is also important to note that all configurations at scale used a 3D domain decomposition, representing the most demanding communication scenario.

\subsection{Device-side timing analysis} 

We carried out a separate set of tests to characterize the device-side execution and the overlap between computation and communication.
To do so, we manually instrumented selected kernels (all versions of Pack and Unpack kernels, as well as Non-bonded kernels) to record kernel start and/or completion time.

The device-side time was read from the \texttt{\%\%globaltimer} register by the first thread in each block and stored in global memory using \texttt{atom.release.gpu.global.cas.b64} (start of the kernel) and \texttt{atom.gpu.global.max.u64} (end of the kernel). The intervals of interest were recorded for each time-step and reported as averages over 200 time-steps to balance granularity and overheads.
In our tests, the instrumentation overhead on end-to-end performance was $<10$\% with smaller inputs and $<1$\% with at least a few hundred thousand atoms per GPU, less than the typical overhead with \texttt{nsys}.
The collected timings were used to compute \textit{Local work} (from the start to the end of the local non-bonded kernel), \textit{Non-local work} (from the start of the first pack to the end of the last unpack), and \textit{Non-overlap} (from the end of the local non-bonded kernel to the end of the last unpack, clamped at zero). The mean \textit{Time per step} was calculated by subtracting the ``Domain Decomposition'' and ``Neighbor Search'' wall-times (CPU tasks performed every 200 steps) from the total run time and averaging over the number of steps.

\subsubsection*{Intra-node}

\begin{figure}
    \centering
    \includegraphics[width=\linewidth]{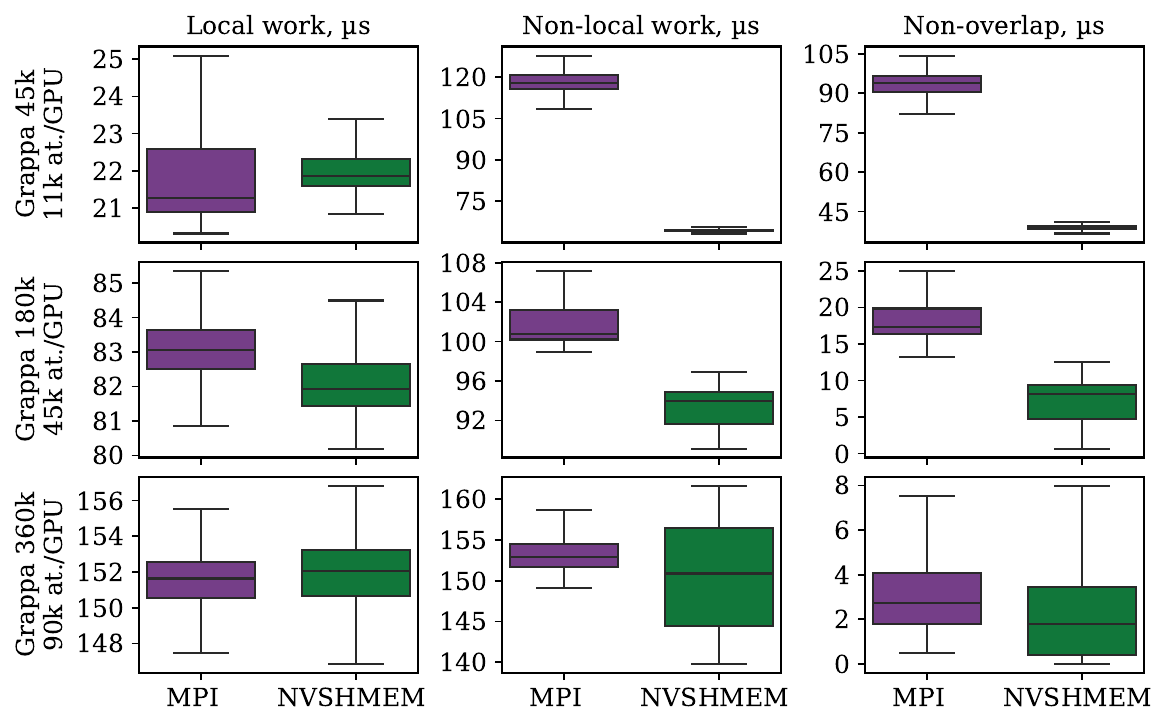}
    \caption{Device-side timing results for intra-node runs, comparing MPI with NVSHMEM. Grappa 45k, 180k, and 360k systems are run on 4 ranks, with 11.25, 45, and 90 thousand atoms per GPU, respectively.}
    \Description{Nine bar charts arranged in a 3x3 grid showing device-side timing measurements (in microseconds) for single-node run. Three rows represent different Grappa system sizes: 45k atoms (11.25k atoms/GPU), 180k atoms (45k atoms/GPU), and 360k atoms (90k atoms/GPU), run on 4 ranks respectively. Three columns show ``Local work'', ``Non-local work', and ``Non-overlap'' timing measurements. Each chart compares MPI and NVSHMEM implementations. For Grappa 45k: ``Local work'' ranges 21-25 μs for both MPI and NVSHMEM, ``Non-local work``: under 75 μs for NVSHMEM and around 120 µs for MPI, ``Non-overlap'': under 45 µs for NVSHMEM and around 90 µs for MPI. For Grappa 90k: ``Local work'' ranges 80-85 μs for both MPI and NVSHMEM, ``Non-local work``: around 92 μs for NVSHMEM and around 102 µs for MPI, ``Non-overlap'': around 8 µs for NVSHMEM and around 18 µs for MPI. For Grappa 180k, both MPI and NVSHMEM show the same behavior, with slightly higher variance for NVSHMEM: ``Local work'': around 152 μs, ``Non-local work``: around 150 μs, ``Non-overlap'': around 2 µs.}
    \label{fig:device-timings-eos-intranode-11k}
\end{figure}

Single-node results for different system sizes, from 45k (11.25k atoms per GPU) to 360k (90k atoms per GPU) are shown in \figref{fig:device-timings-eos-intranode-11k}. These tests are run on 4 GPUs with NVLink and use 1D DD.
The local work duration grows nearly linearly with the increase in system size (1.7-2.0 ns/atom) and shows little dependence on the communication method.
For small systems, in the CPU-bound regime, the advantage of NVSHMEM is the largest (64 µs vs 116 µs), as it replaces multiple operations (kernel launch, device-host synchronization, MPI send, MPI receive) with a single one. Non-local work takes much longer than local, and it is the rate-limiting part.
As the system size increases and the simulation becomes less latency-bound, we even see, in the case of MPI, the reduction of the non-local work time from 116 µs to 101 µs when going from 11.25k to 45k atoms/GPU, despite the amount of compute work and communication volume (but not the number of pulses) increasing. NVSHMEM performance predictably decreases to 94 µs, but it is still better than MPI and able to nearly fully overlap local and non-local work.
Further increasing the system size to 90k atoms/GPU, the local and non-local work durations become nearly equal ($\sim$152 µs), enabling near-perfect overlap. As a result, the non-local compute becomes substantial enough to render the performance difference between the two communication approaches negligible.

\subsubsection*{Multi-node}

As we scale GROMACS across more GPUs, the DD automatically switches from 1D to 2D and finally to 3D, based on the simulation box shape and number of ranks. Higher-dimensional decompositions increase the number of communication pulses, cases where we expect our NVSHMEM approach to show a larger advantage over MPI. This scaling also introduces inter-node communication over InfiniBand. To investigate this, we compare two regimes: 11.25k and 90k atoms per GPU.
The tests were run on 8, 16, and 32 GPUs (2, 4, and 8 nodes), which correspond to the 1D, 2D, and 3D decompositions, respectively.


\begin{figure}
    \centering
    \includegraphics[width=\linewidth]{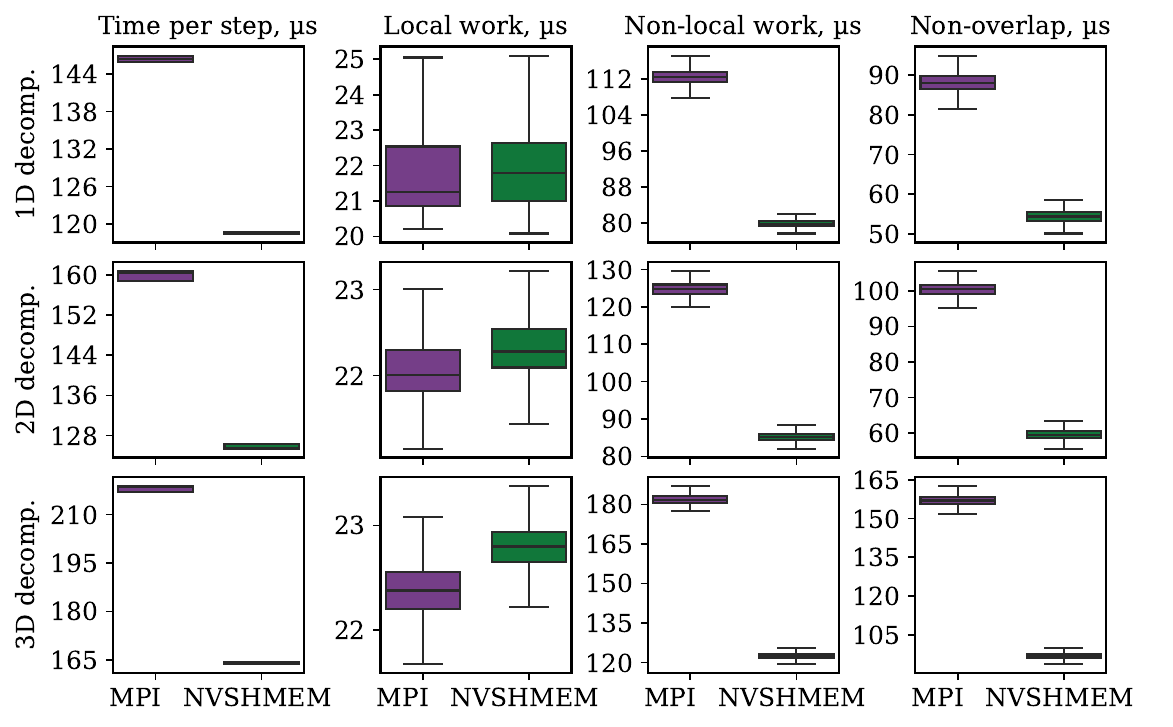}
    \caption{Device-side timing results for multi-node runs with 11.25k atoms per GPU. Grappa 90k, 180k, and 360k systems are run on 8, 16, and 32 ranks.}
    \Description{Twelve bar charts arranged in a 3x4 grid showing device-side timing measurements (in microseconds). The rows correspond to 1D, 2D, and 3D decompositions, respectively. Each column correspond to a different metric: ``Time per step'', ``Local work'', ``Non-local work``, and ``Non-overlap''. Each bar chart compares MPI and NVSHMEM results for given decomposition and metric. The ``Local work'' is consistently around 22 µs for all cases. NVSHMEM has consistently lower ``Time per step``, ``Non-local work``, and ``Non-overlap`` than MPI, with the difference increasing with increasing decomposition dimensionality.}
    \label{fig:device-timings-eos-11k}
\end{figure}

Fig.~\ref{fig:device-timings-eos-11k} shows results with 11.25k atoms/GPU. The total duration of local work is, like in the intra-node case, $\sim$22 µs, and the non-local work takes at least 80 µs. The total time per step is limited by the non-local work, other tasks contribute 30-40 µs regardless of the DD setup.
Going from 1D to 2D changes the non-local work duration by under 11\% despite doubling the number of communication pulses.
Going from 2D to 3D increases the non-local time by $\sim$45\%, consistent with the $1.5\times$ increase in the number of pulses.
Even at this size, we see a minor (sub-µs) effect of NVSHMEM on local work: since NVSHMEM uses SM resources for communications, overlapping local work is slowed down.

\begin{figure}
    \centering
    \includegraphics[width=\linewidth]{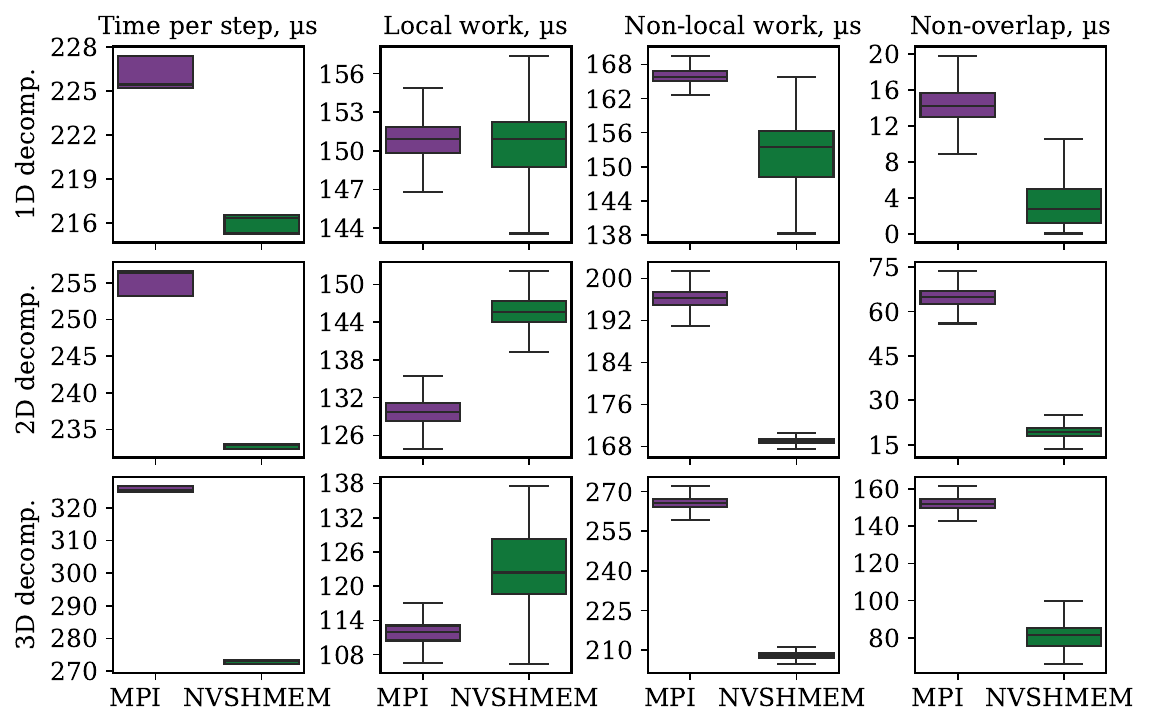}
    \caption{Device-side timing results for multi-node runs with 90k atoms per GPU. Grappa 720k, 1440k, and 2880k systems are run on 8, 16, and 32 ranks.}
    \Description{Twelve bar charts arranged in a 3x4 grid showing device-side timing measurements (in microseconds). The rows correspond to 1D, 2D, and 3D decompositions, respectively. Each column correspond to a different metric: ``Time per step'', ``Local work'', ``Non-local work``, and ``Non-overlap''. Each bar chart compares MPI and NVSHMEM results for given decomposition and metric. The ``Local work'' is identical for MPI and NVSHMEM when using 1D decomposition; for 2D and 3D cases, MPI has lower timings, by around 10 µs. NVSHMEM has consistently lower ``Time per step``, ``Non-local work``, and ``Non-overlap`` than MPI, with the difference increasing with increasing decomposition dimensionality.}
    \label{fig:device-timings-eos-90k}
\end{figure}

Fig.~\ref{fig:device-timings-eos-90k} shows results with 90k atoms per GPU.
In the 1D case, the local and the non-local work duration are close (151 µs vs. 153-165 µs).
Although a small portion of the non-local work remains non-overlapped, it is much smaller and it is completely eliminated in some cases with NVSHMEM
(the non-local work is fully overlapped with local work). Thus, for the 1D case, the communication method has a minimal impact on the total step time ($\sim$10 µs).
When we switch to 2D decomposition, the difference in non-local work duration between MPI and NVSHMEM increases to $\sim$28 µs. Resource sharing with NVSHMEM communication slows down the local kernel by $\sim$16 µs, but this kernel is not on the critical path, and its slowdown has little impact: the total time per step is $\sim$24 µs shorter with NVSHMEM than with MPI, thanks to the reduced non-local time.
This trend continues when going to 3D decomposition, with NVSHMEM being faster than MPI by 50-60 µs in both non-local work and the total time per step, despite slowing down local work by $\sim$10 µs.

Extrapolating to larger number of atoms per GPU, with low-dimensionality decomposition when the local work ends up on the critical path, the downsides of the NVSHMEM approach (slower local work) can start to outweigh its benefits (faster non-local work), and MPI becomes the slightly faster option, consistent with our observations with 23040k atoms on 2 and 4 nodes ($\ge$ 1440k atoms/GPU); see Fig.~\ref{fig:strong-scaling-comparison}.


\section{Conclusions and Future work}

Post-exascale heterogeneous HPC architectures pose major challenges to latency-sensitive strong-scaling application like MD, while established programming models struggle to keep up with the evolution of hardware, and standards like MPI show limitations to efficiently utilize heterogeneous systems. A wide range of applications, including MD and ultimately most codes that strong-scale to sufficiently far, require tighter integration of communication control paths, lower latencies, and fine-grained compute and communication overlap. While there is ongoing work for better GPU support in the MPI standard and in early implementations \cite{bridges_understanding_2024, temucin_design_2024,intel_mpi_docs,namashivayam_exploring_2023}, there are no mature solutions available. PGAS-style one-sided communication libraries like NVSHMEM are currently best suited for this, partly due to a simpler and lower-level API, but also to extensive vendor extensions and optimizations that go beyond the standard and allow GPU-initiated communication and thread-collaborative warp/thread-block level APIs. We harness these to improve the scalability of GROMACS MD simulations through a novel GPU-initiated domain decomposition halo exchange algorithm. Our GPU-resident halo exchange formulation combines data packing with data movement and receiver notification.
Importantly, we fuse coupled communication phases across all dimensions and pulses by separating dependent and independent data for each pulse. Unlike CPU-initiated formulations that suffer from control-path latencies and synchronization overhead due to data dependencies inherent to the neutral-territory approach, our fused kernel design maximizes concurrency and enables fine-grained overlap by harnessing the GPU hardware latency hiding. The NVSHMEM implementation targets both inter- and intra-node scenarios, employing extensive low-level optimizations including intra-node signaling and TMA engine-based data movement.

The resulting implementation demonstrates superior application performance and improved parallel efficiency compared to the optimized GROMACS GPU-aware MPI halo exchange across a wide range of the studied scaling regimes across NVLink and InfiniBand+NVLink. 

The major speedup across NVLink both intra- and multi-node highlights the ability of NVSHMEM-based GPU-initiated communication to exploit high-performance interconnects.
Using NVSHMEM is not without drawbacks since using GPU SMs for synchronization and communication can lead to resource competition with overlapping compute work. This can have performance tradeoffs, but in our case the impact on GROMACS performance is small and it only negates the benefits compared to MPI on small node-counts with large input sizes. To reduce the GPU SM time spent waiting due to imbalance, our workaround syncs up PEs with a CPU-based synchronization which reduces the resource competition-related overhead. As the resulting implementation is no longer fully GPU-resident, we plan to design a GPU-resident solution for this.
The requirement related to symmetric allocation can pose significant programming and application maintainability challenges as we have seen with the GROMACS rank-specialization. We hope that this drawback can be resolved with a team-based allocation extension in NVSHMEM.

Opportunities remain to further fuse the halo exchange kernels with the non-local compute work to expose more intra-kernel compute-communication overlap. We also plan use the GPU-ini\-ti\-at\-ed communication approaches and optimizations employed here to redesign the rest of the communication in GROMACS, notably the communication of coordinates and forces to and from the PME tasks which will be key to fully unlock the scalability potential of important GROMACS workloads.

While our evaluation focused on multi-node InfiniBand, we observed similar trends on HPE Slingshot 11 at smaller scales. A more thorough assessment on Slingshot and multi-node NVLink remains to be explored in future work.

Our results and analysis highlight the profound benefits of GPU-initiated communication for latency reduction and improved strong scaling in a broad range of HPC applications. Halo exchange algorithms similar to the one used in MD are common in domains such as CFD and astrophysics.
Therefore, the algorithms and optimizations presented in this work are highly reusable and can benefit other HPC codes aiming to improve performance on current and future heterogeneous architectures.

\begin{acks}
This work was
\grantsponsor{ganana}{funded by the European Union under the GANANA project}{https://doi.org/10.3030/101196247} (Grant Agreement No.~\grantnum{ganana}{101196247})
and supported by the Swedish e-Science Research Center.
We thank Berk Hess and Mark Abraham for insightful discussions and feedback on the initial halo exchange design and code review during its upstreaming.
We thank Akhil Langer for providing initial guidance on NVSHMEM and Tam\'as B\'ela Feh\'er for his thoughtful manuscript review and support.
For hardware access during development, we thank
FZ J\"ulich (projects exalab/slbio),
CSC (project 2003389),
and CSCS (project g174).
\end{acks}

\bibliographystyle{ACM-Reference-Format}
\bibliography{biblio}

\appendix

\pagenumbering{gobble} 


\twocolumn[%
{\begin{center}
\Huge
Appendix: Artifact Description/Artifact Evaluation
\end{center}}
]

\renewcommand{\artexpl}[1]{}


\appendixAD

\section{Overview of Contributions and Artifacts}

\subsection{Paper's Main Contributions}

\artsampl{
\begin{description}
\item[$C_1$] NVSHMEM-based GPU-initiated halo exchange implementation in GROMACS.
\item[$C_2$] Critical path schedule optimizations.
\item[$C_3$] NVSHMEM proxy thread affinity setting prototype implementation in GROMACS.
\item[$C_4$] Intra- and inter-node performance benchmarks and strong scaling analysis.
\item[$C_5$] GPU device timing-based kernel execution and critical path analysis.
\end{description}
}

\subsection{Computational Artifacts}

\artsampl{
\begin{description}
\item[$A_1$] https://doi.org/10.5281/zenodo.16572713
\item[$A_2$] https://doi.org/10.5281/zenodo.17062607
\end{description}
}

\begin{center}
\begin{tabular}{rll}
\toprule
Artifact ID  &  Contributions &  Related \\
             &  Supported     &  Paper Elements \\
\midrule
$A_1$   &  $C_1$-$C_5$ & Figures 3-8 \\
\midrule
$A_2$   &  $C_1$-$C_5$ & Figures 3-8 \\
\bottomrule
\end{tabular}
\end{center}

\section{Artifact Identification}

\artexpl{
Provide the following six subsections for each computational artifact $A_i$.
}

\newartifact

\artrel

This artifact contains the GROMACS 2025-based source code which implements the algorithms presented as well as the supporting code for the GPU-kernel timing analysis.

\artexp

The NVSHMEM-based GPU-initiated halo exchange is expected to outperform the MPI-based halo exchange both intra- and inter-node scenarios for the range of scaling regimes presented in the manuscript on architectures similar to those benchmarked (NVIDIA Hopper GPUs with NVLink intra-node and multi-rail InfiniBand NDR inter-node interconnects).

\arttime

Multi-node strong scaling performance benchmarks (Figs.~4-5) are expected to take 1-22 minutes per data point.
Intra-node performance benchmarks (Fig.~3) are expected to take 1-5 minutes per data point.
Note that longer than 5 minute runtime should only be expected for the large inputs on very low node counts. 

\artin

\artinpart{Hardware}

For intra-node and InfiniBand multi-node runs (Figs.~3,5): 
4 H100 GPUs per node with 4th generation NVLink (8 GPUs for intra-node); 2 sockets of Intel Sapphire Rapids 56 core CPUs per node. For multi-node, additionally multi-rail InfiniBand with 400~Gb/s NDR inter-node interconnects. 

For MNNVL scaling (Fig.~4): NVL72 based multi-node NVLink in a 36x2 configuration with 4 GB200 GPUs and 2 NVIDIA Grace 72 core CPUs per node. 

\artinpart{Software}

\artexpl{
Introduce all required software packages, including the computational artifact. For each software package, specify the version and provide the URL.
}
\begin{itemize}
    \item gromacs-nvshmem-halo-exchange.tar.gz: extending the upstream 2025.2 GROMACS release \\(\url{https://gitlab.com/gromacs/gromacs}, \\git tag ae1179d874e9c52e402e3f67df5e77ee5ef032ed) with a new, optimized NVSHMEM implementation of GPU-initiated halo exchange (corresponding to \\git commit ad76c2c88c81ba4e3c70791f4c017a3004d16e2a).
    \item device-timings-patch.tar.gz: the patch for the modified GROMACS version to add the device-side time measurement, corresponding to (\url{https://gitlab.com/gromacs/gromacs}, \\git commit 45ff933eccd310e8e24127a629e102d80ae68c0d)
    \item NVSHMEM 3.1.7 (3.2.5 for MNNVL runs) (with IBRC transport): \url{https://developer.nvidia.com/nvshmem-archive}.
    \item Open MPI 4.1.7rc1 (5.0.6 for MNNVL runs): \url{https://www.open-mpi.org/software/ompi/v4.1/}
    \item UCX 1.19.0: \url{https://github.com/openucx/ucx/releases}
\end{itemize}

\artinpart{Datasets / Inputs}

\artexpl{
Describe the datasets required by the artifact. Indicate whether the datasets can be generated, including instructions, or if they are available for download, providing the corresponding URL.
}

All runs use the `grappa' set of benchmark inputs, provided in $A_2$, \verb|grappa-45k-23M.tar.gz|. There is one directory per input size ($45000$ -- $23$ million atoms). 

These inputs require pre-processing before running. Extract the tarball and edit \verb|prepare_inputs.sh|. Set the environment variable \verb|GMX| to point to the GROMACS binary (see `Installation and Deployment'). Set \verb|INPUT_DIR| to point to the root of the grappa inputs directory, then run \verb|prepare_inputs.sh|. This will generate one .tpr file in each input directory. 

\artinpart{Installation and Deployment}

\artexpl{
Detail the requirements for compiling, deploying, and executing the experiments, including necessary compilers and their versions.
}

For intra-node and InfiniBand multi-node runs, GROMACS was compiled using the following environment: Ubuntu 22.04, CUDA Toolkit 12.6.3, CUDA Driver 535.129.03, GCC 13.1.0, Open MPI 4.1.7rc1, UCX 1.19.0, NVSHMEM 3.1.7 (with IBRC transport), and CMake 3.30.3.

To compile GROMACS use the \verb|build_gromacs.sh| script (found in \verb|eos_inter_node_grappa_logs.tar.gz| in $A_2$) with:
\begin{itemize}
    \item `mpi' argument for the MPI build which will generate the application binary \verb|build_mpi/bin/gmx_mpi|.
    \item `nvshmem' argument for the NVHSMEM build which will generate the application binary \verb|build_nvshmem/bin/gmx_mpi|. 
\end{itemize}
Compilation scripts should be run from the root of the GROMACS source tree. 

For intra-node runs, ensure NVSHMEM shared libraries are discoverable at runtime. If NVSHMEM is not installed system-wide, export the library path before running, e.g.~ \verb|export|\\\verb|LD_LIBRARY_PATH=/path/to/nvshmem/lib:$LD_LIBRARY_PATH|.\\The inter-node SLURM wrapper scripts set this automatically.

For the multi-node NVLink (MNNVL) runs GROMACS was compiled using the following environment: Ubuntu 24.04.2 LTS, CUDA Toolkit 12.8.61, CUDA Driver 570.172.08, GCC 13.3.0, Open MPI 5.0.6, UCX 1.19.0, NVSHMEM 3.2.5 (with IBRC transport), and CMake 3.31.6.

To compile, run \verb|mnnvl_cluster_nvshmem_logs/build_mnnvl.sh| from \verb|mnnvl_cluster_nvshmem_logs.tar.gz| in $A_2$ with \mbox{`nvshmem'} argument. This will generate the \verb|build_nvshmem/bin/gmx_mpi| application binary. 

\artcomp

\artexpl{
Provide an abstract description of the experiment workflow of the artifact. It is important to identify the main tasks (processes) and how they depend on each other.

A workflow may consist of three tasks: $T_1, T_2$, and $T_3$. The task $T_1$ may generate a specific dataset. This dataset is then used as input by a computational task $T_2$, and the output of $T_2$ is processed by another task $T_3$, which produces the final results (e.g., plots, tables, etc.). State the individual tasks $T_i$ and provide their dependencies, e.g., $T_1 \rightarrow T_2 \rightarrow T_3$.

Provide details on the experimental parameters. How and why were parameters set to a specific value (if relevant for the reproduction of an artifact), e.g., size of dataset, number of data points, input sizes, etc. Additionally, include details on statistical parameters, like the number of repetitions.
}

1) Build the GROMACS \verb|gmx_mpi| binary using \verb|build_gromacs.sh| as described in `Installation and Deployment'.

2) Prepare inputs as described in `Datasets / Inputs' 

3) Generate the raw data for the intra-node case (Fig.~3). Run GROMACS on 4 and 8 GPUs with MPI and NVSHMEM, generating a log file for each run. 

To do this, untar \verb|mpi_tmpi_nvshmem_intranode_eos.tar.gz| from $A_2$ and run \verb|run_md_tests_intra_node_mpi.sh| for MPI runs and \verb|run_md_tests_intra_node_nvshmem.sh| for NVSHMEM runs, updating the path to the grappa inputs root directory (\verb|TPR_DIR|) and GROMACS build directory (\verb|MDRUN_DIR|). This will generate a separate log file for 4 and 8 GPUs for 4 different input sizes in the directories:

\texttt{mdrun\_logs/nvshmem\_timing\_runs\_nstlist\_200\_intranode} for NVSHMEM intra-node results and\
\\texttt{mdrun\_logs/mpi\_timing\_runs\_nstlist\_200\_intranode} MPI intra-node results.

For intra-node NVSHMEM runs, if you encounter dynamic loader errors related to NVSHMEM (e.g., missing \texttt{libnvshmem*.so}), set \verb|LD_LIBRARY_PATH| to include the NVSHMEM \verb|lib| directory as described in `Installation and Deployment'.

4) Generate the raw data for the multi-node IB case (Fig.~5) Run GROMACS on a range of process counts and input sizes with MPI and with NVSHMEM, generating one log file for each run. 

To do this, untar \verb|eos_inter_node_grappa_logs.tar.gz| from $A_2$ and run \verb|run_md_tests_inter_node_slurm_mpi.sh| for MPI and \verb|run_md_tests_inter_node_slurm_nvshmem.sh| for \\ NVSHMEM, updating the path to the root directory of the grappa inputs (\verb|TPR_DIR|) and GROMACS build directory (\verb|MDRUN_DIR|) accordingly. NVSHMEM use is enabled at runtime with the \\\verb|GMX_ENABLE_NVSHMEM=1|. Note this will generate and submit one job script per input size per node count, as defined in the \\\verb|dir_numbers| and \verb|ntasks_list| arrays. This will generate a separate log file for each job at\\\verb|mdrun_logs/inter_node_runs_nstlist_200_simd_auto|.

Note if a node with a different number of cores/sockets than described in `Hardware' is used, care should be taken to adjust `wrapper.sh' and number of OpenMP threads such that threads in a single rank do not span socket boundaries and each rank uses fewer threads than the number of cores available to that rank.

We expect best results with rank-level pinning, however this may depend on cluster. The following can be used to benchmark thread-level pinning. 

Untar \verb|nvshmem_proxy_pinning.tar.gz| and run \\ \verb|job_run_grappa_pinToCore.sh|, updating the path to the grappa input root directory (\verb|INPUT_DIR|) and location of the GROMACS binary (\verb|GMX|) This will generate one log file for the input size specified by \verb|GRAPPA_SIZE| and number of GPUs specified by\\\verb|--ntasks-per-node| in the at \verb|log_files_grappa_*|. This script will run with thread-level pinning with the last thread reserved for the NVSHMEM proxy thread, enabled with\\\verb|GMX_NVSHMEM_RESERVE_THREAD=1|. This can be compared to the results with rank-level pinning above. 

5) Generate the raw data for the MNNVL case (Fig.~4). Run GROMACS on three input sizes for a range of node counts. 

Untar \verb|mnnvl_cluster_nvshmem_logs.tar.gz| and run \\ \verb|run_md_tests_mnnvl.sh| and update the path to the root directory of the grappa inputs and the location of the GROMACS binary.

6) Post-processing of log files to generate figures will be described in the section for Artifact 2. 

\artout

The following set of output directories should be generated:

Intra-node case:
\begin{itemize}
    \item\texttt{mdrun\_logs/}\\ \texttt{nvshmem\_timing\_runs\_nstlist\_200\_intranode}
    \item\texttt{mdrun\_logs/mpi\_timing\_runs\_nstlist\_200\_intranode} 
\end{itemize}
Multi-node case:
\begin{itemize}
    \item\verb|mdrun_logs/inter_node_runs_nstlist_200_simd_auto|
\end{itemize}
NVSHMEM proxy thread affinity experiments:
\begin{itemize}
    \item \verb|log_files_grappa_*|
\end{itemize}

\newartifact

\artrel

\artexpl{
    Briefly explain the relationship between the artifact and contributions.
}

This artifact contains post-processing scripts to generate figures from the raw data generated by running artifact $A_1$ as well as input data-set; for reference the benchmark logs/outputs  are also included.

\artexp

\artexpl{
Provide a higher level description of what outcome to expect from the corresponding experiments. Provide an explanation of how the results substantiate the main contributions.
}

Post processing of raw data to generate Figures 3-8. 

\arttime

\artexpl{
Estimate the time required to reproduce the artifact, providing separate estimates for the individual steps: Artifact Setup, Artifact Execution, and Artifact Analysis.
}

Run time of analysis scripts is under one minute.

\artin

\artinpart{Hardware}

\artexpl{
Specify the hardware requirements and dependencies (e.g., a specific interconnect or GPU type is required).
}

No special hardware requirements for the data analysis scripts.

\artinpart{Software}

\artexpl{
Introduce all required software packages, including the computational artifact. For each software package, specify the version and provide the URL.
}

Device-side timing analysis scripts were run on Ubuntu 24.04, using Python~3.12.3. Required Python packages:
\begin{itemize}
    \item Pandas~2.1.4 (https://pypi.org/project/pandas/2.1.4/)
    \item Seaborn~0.13.2 (https://pypi.org/project/seaborn/0.13.2/)
    \item Matplotlib~3.6.3 (https://pypi.org/project/matplotlib/3.6.3/)
    \item Tqdm~4.66.2 (https://pypi.org/project/tqdm/4.66.2/)
\end{itemize}

\artinpart{Datasets / Inputs}

\artexpl{
Describe the datasets required by the artifact. Indicate whether the datasets can be generated, including instructions, or if they are available for download, providing the corresponding URL.
}

This artifact includes all inputs needed for post-processing and plotting:
\begin{itemize}
    \item \verb|grappa-45k-23M.tar.gz| (benchmark set, for reference)
    \item \verb|mpi_tmpi_nvshmem_intranode_eos.tar.gz| (intra-node logs, build and run scripts)
    \item \verb|eos_inter_node_grappa_logs.tar.gz| (inter-node logs, build and run scripts)
    \item \verb|nvshmem_proxy_pinning.tar.gz| (proxy pinning logs and scripts)
    \item \verb|mnnvl_cluster_nvshmem_logs.tar.gz| (MNNVL logs, build and run scripts)
    \item \verb|device-timings.tar.gz| (device-side timing inputs and scripts)
\end{itemize}
The plotting scripts consume the provided logs/outputs directly. If you regenerate logs yourself using $A_1$, point the parsers to the corresponding directories produced by $A_1$.

\artinpart{Installation and Deployment}

\artexpl{
Detail the requirements for compiling, deploying, and executing the experiments, including necessary compilers and their versions.
}

Installing dependencies for device-side timing analysis scripts can be done as follows:
\begin{itemize}
    \item If running on Ubuntu 24.04: \texttt{sudo apt install python3.12 python3-pandas python3-matplotlib python3-seaborn python3-tqdm}
    \item If using Pip: \texttt{pip install -r requirements.txt}
\end{itemize}
After installing Python packages, unpack the $A_2$ archives into a working directory (each \verb|*.tar.gz| creates its own subdirectory). Ensure the expected log directories and \verb|eos/| inputs exist before running the plotting scripts below.

\artcomp

\artexpl{
Provide an abstract description of the experiment workflow of the artifact. It is important to identify the main tasks (processes) and how they depend on each other.

A workflow may consist of three tasks: $T_1, T_2$, and $T_3$. The task $T_1$ may generate a specific dataset. This dataset is then used as input by a computational task $T_2$, and the output of $T_2$ is processed by another task $T_3$, which produces the final results (e.g., plots, tables, etc.). State the individual tasks $T_i$ and provide their dependencies, e.g., $T_1 \rightarrow T_2 \rightarrow T_3$.

Provide details on the experimental parameters. How and why were parameters set to a specific value (if relevant for the reproduction of an artifact), e.g., size of dataset, number of data points, input sizes, etc. Additionally, include details on statistical parameters, like the number of repetitions.
}

To generate Figs.~3-5 (strong scaling performance comparison charts), run \texttt{python3 generate\_performance\_charts.py}. The output figures in PDF format are produced in the current directory.

Device-side timing analysis scripts require the input data (logs and \texttt{stdout} files produced by the instrumented version of GROMACS) placed in the \texttt{eos} directory.

To generate Fig.~6 (``Device-side timing results for intra-node run''), run\\\texttt{python3 plot\_combined\_performance\_eos\_intranode.py}. The output figure in the PDF format is produced in the current directory.

To generate Figs.~7-8 (``Device-side timing results for multi-node runs''), run\\\texttt{python3 plot\_combined\_performance\_eos.py}. The output figures in the PDF format are produced in the current directory.

\artout

Figures 3-8 (in PDF format)

\newpage
\appendixAE

\arteval{1}
\artin

Install CUDA, MPI, UCX, NVSHMEM as in \S\,Artifact Identification (A\_1). Build \verb|gmx_mpi| as described (MPI and NVSHMEM variants). Ensure SLURM access for multi-node; verify NVLink/InfiniBand connectivity with cluster diagnostics.

\artcomp

\noindent Workflow overview (tasks and dependencies).
\begin{description}
\item[Task1] Build environment and binaries (\verb|gmx_mpi|, MPI and NVSHMEM variants)
\begin{itemize}
\item \textbf{Input}: source tree ($A_1$), toolchain (CUDA, MPI, UCX, NVSHMEM), \texttt{eos\_inter\_node\_grappa\_logs.tar.gz} ($A_2$)
\item \textbf{Steps}: run \verb|build_gromacs.sh mpi| and \\ \verb|build_gromacs.sh nvshmem|; verify \verb|gmx_mpi| exists
\item \textbf{Output}: \verb|build_mpi/bin/gmx_mpi|, \\\verb|build_nvshmem/bin/gmx_mpi|
\end{itemize}
\item[Task2] Preprocess benchmark inputs
\begin{itemize}
\item \textbf{Input}: \verb|grappa-45k-23M.tar.gz| ($A_2$)
\item \textbf{Steps}: set \verb|GMX| to built binary; run \verb|prepare_inputs.sh| to generate one \verb|.tpr| per system size
\item \textbf{Output}: per-size \verb|.tpr| files under the input directories
\end{itemize}
\item[Task3] Intra-node experiments (Fig.~3). Depends on: Task1, Task2
\begin{itemize}
\item \textbf{Input}: \verb|.tpr| files; 4 or 8 GPUs/node
\item \textbf{Steps}: run run\_md\_tests\_intra\_node\_mpi.sh (MPI) and run\_md\_tests\_intra\_node\_nvshmem.sh| (NVSHMEM); set \verb|TPR_DIR| and \verb|MDRUN_DIR| appropriately; ensure CPU/GPU binding is consistent across runs. for intra-node NVSHMEM runs, export \verb|LD_LIBRARY_PATH| to include the NVSHMEM \verb|lib| directory if your environment does not already set it (see A\_1 Installation and Deployment)
\item \textbf{Output}: under the \verb|mdrun_logs/| directory, in\\\verb|mpi_timing_runs_nstlist_200_intranode| and\\\verb|nvshmem_timing_runs_nstlist_200_intranode| subdirectories.
\end{itemize}
\item[Task4] Inter-node experiments over InfiniBand (Fig.~5). Depends on: Task1, Task2
\begin{itemize}
\item \textbf{Input}: \verb|.tpr| files; SLURM access; node counts per \verb|ntasks_list|
\item \textbf{Steps}: run \verb|run_md_tests_inter_node_slurm_mpi.sh| (MPI) and \verb|run_md_tests_inter_node_slurm_nvshmem.sh| (NVSHMEM); set \verb|TPR_DIR| and \verb|MDRUN_DIR|; the scripts generate and submit one job per (size, node count)
\item \textbf{Output}: logs in\\\verb|mdrun_logs/inter_node_runs_nstlist_200_simd_auto|
\end{itemize}
\item[Task5] Multi-node NVLink (MNNVL) experiments (Fig.~4). Depends on: Task1, Task2
\begin{itemize}
\item \textbf{Input}: \verb|.tpr| files; GB200 NVL72 cluster access
\item \textbf{Steps}: build with \verb|build_mnnvl.sh|; \\run \verb|run_md_tests_mnnvl.sh| after setting input and binary paths
\item \textbf{Output}: logs in the MNNVL run directory (as provided in $A_2$)
\end{itemize}
\item[Task6] Collect and validate outputs. Depends on: Task3, Task4, Task5
\begin{itemize}
\item \textbf{Steps}: check each expected run produced a log; extract \emph{Performance (ns/day)} entries; summarize into CSVs; if multiple repetitions exist, report medians.
\item \textbf{Output}: complete set of logs and summary CSVs used by $A_2$.
\end{itemize}
\end{description}
\noindent Dependencies: Task1 $\rightarrow$ Task2 $\rightarrow$ (Task3, Task4, Task5) $\rightarrow$ Task6.\\
\noindent Key parameters: system sizes (45k-23M atoms), GPU counts/node counts, \verb|nstlist=200|, repetitions per script defaults. Maintain identical binding and environment across MPI vs. NVSHMEM for fairness.

\artout

Expected: directories and logs as listed above; reproduced performance trends for Figs.~3--5. 

\begin{description}
\item[Expected results] Strong-scaling performance curves that replicate Figs.~3--5: for each system size and scale (GPUs/nodes), NVSHMEM achieves performance that is at least on par with, and generally higher than, MPI, with the performance gap typically widening at larger scales. 
\item[Methodology for evaluation] Parse each run's mdrun log and extract the final \emph{Performance (ns/day)}. For repeated runs, take the median. Construct per-figure CSVs and compute the speedup \(S = \frac{\text{NVSHMEM}}{\text{MPI}}\) (\(S>1\) indicates NVSHMEM is faster) for matching configurations. Recreate the line/bar plots for Figs.~3--5 and verify: (i) strong-scaling trends, (ii) NVSHMEM curves lie at or above MPI for most of the points, and (iii) relative ranking and crossovers match the paper. 
\item[Relation to contributions] $C_1$ (NVSHMEM GPU-initiated halo exchange) is corroborated by higher NVSHMEM performance vs. MPI across scales. $C_2$ (critical-path schedule optimizations) is reflected by the growing advantage at larger scales, consistent with improved overlap and reduced communication on the critical path. $C_3$ (proxy thread affinity) is supported by the absence of a detriment when using rank-level pinning only. $C_4$ (benchmarks and strong scaling analysis) is reproduced by intra-/inter-node trends and measured efficiencies. $C_5$ is further corroborated in the analysis of $A_2$ (device-side timing), which characterizes where communication and computation overlap is better and runtime is improved.
\item[Figure correspondence] The reproduced Figs.~3--5 generated from your runs should match the generalizable outcomes in the paper: the relative ordering of methods, curvature/slopes, and the scale at which the NVSHMEM advantage becomes pronounced.
\end{description}

\arteval{2}
\artin

Python 3.10.12 with pandas, seaborn, matplotlib, tqdm. No special hardware.

Required $A_2$ bundles (\url{https://zenodo.org/records/16622042}):
\begin{itemize}
    \item \verb|mpi_tmpi_nvshmem_intranode_eos.tar.gz|
    \item \verb|eos_inter_node_grappa_logs.tar.gz|
    \item \verb|mnnvl_cluster_nvshmem_logs.tar.gz|
    \item \verb|device-timings.tar.gz|
\end{itemize}
\artcomp

Untar the archives from $A_2$:
\begin{itemize}
    \item Fig.~3: \texttt{mpi\_tmpi\_nvshmem\_intranode\_eos.tar.gz}
    \item Fig.~4: \texttt{mnnvl\_cluster\_nvshmem\_logs.tar.gz}
    \item Fig.~5: \texttt{eos\_inter\_node\_grappa\_logs.tar.gz}
    \item Figs.~6--8: \texttt{device-timings.tar.gz}
\end{itemize}

Workflow overview (tasks and dependencies).
\begin{description}
\item[Task1] Verify and unpack A\_2 bundles
\begin{itemize}
\item \textbf{Input}: A\_2 archive files (see list above)
\item \textbf{Steps}: confirm presence; unpack each \verb|*.tar.gz| into a working area; note created directories
\item \textbf{Output}: unpacked directories containing logs, helper scripts, and plotting utilities
\end{itemize}
\item[Task2] Stage inputs for analysis
\begin{itemize}
\item \textbf{Input}: unpacked directories from Task1
\item \textbf{Steps}: place device timing inputs (from\\\verb|device-timings.tar.gz| and instrumented outputs) under \verb|eos/| as expected by the scripts; ensure intra-node and inter-node logs are available from\\\verb|mpi_tmpi_nvshmem_intranode_eos.tar.gz| and\\\verb|eos_inter_node_grappa_logs.tar.gz|; keep MNNVL logs available from \\\verb|mnnvl_cluster_nvshmem_logs.tar.gz|
\item \textbf{Output}: staged directory layout compatible with the provided Python scripts
\end{itemize}
\item[Task3] Parse logs to structured data
\begin{itemize}
\item \textbf{Input}: mdrun logs (stdout) and device-timing outputs
\item \textbf{Steps}: Extract \emph{Performance (ns/day)}, GPU/node configuration; write to CSV files using the \\ extract\_intranode\_performance.py (used for Fig. 3), extract\_internode\_performance.py  (used for Fig. 5). For fig .4 (MNNVL) results we manually create the CSV in the format  \texttt{performance\_data\_ptyche\_grappa\_rf.csv}.
\item \textbf{Output}: CSVs in the working directory
\end{itemize}
\item[Task4] Generate figures
\begin{itemize}
\item \textbf{Input}: CSVs from Task3
\item \textbf{Steps}: run \verb|generate_performance_charts.py| (Figs.~3--5), \verb|plot_combined_performance_eos_intranode.py| (Fig. 6), \verb|plot_combined_performance_eos.py| (Figs.~7--8)
\item \textbf{Output}: PDFs for Figs.~3--8
\end{itemize}
\item[Task5] Sanity checks. Depends on: Task4.
\begin{itemize}
\item \textbf{Steps}: visually verify relative ranking/trends match the paper; spot-check a few values against CSVs
\item \textbf{Output}: confirmation that regenerated figures are consistent
\end{itemize}
\end{description}
\noindent Dependencies: \; Task1 $\rightarrow$ Task2 $\rightarrow$ Task3 $\rightarrow$ Task4 $\rightarrow$ Task5.
Replace the log files with your results, preserving file names and location.
Run the analysis scripts to regenerate figures.

\artout

PDFs for Figs.~3--8 regenerated from your logs.

\begin{description}
\item[Expected results] Reconstructed figures that visually match the paper's Figs.~3--8: (i) strong-scaling performance comparisons (NVSHMEM \(\geq\) MPI) for intra- and inter-node cases, (ii) MNNVL scaling with consistent trends, and (iii) device-side timing figures showing reduced communication stalls and improved overlap on the critical path for NVSHMEM.
\item[Methodology for evaluation] Use the $A_2$ bundles (\url{https://zenodo.org/records/16622042}). Unpack and stage logs and timing data as described above, then run the provided scripts to regenerate figures. Validate that the regenerated plots preserve the same relative ranking and trends reported in the paper. For device-side timing (Figs.~6--8), confirm that the stacked/segmented timelines and kernel groupings are consistent with the schedule optimizations.
\item[Relation to contributions] $C_1$/$C_4$ are reflected in the reproduced performance figures (Figs.~3--5). $C_2$/$C_5$ are supported by device-side timing breakdowns (Figs.~6--8) that attribute performance gains to less impact of communication on the critical path and improved overlap.
\item[Figure correspondence] The regenerated Figs.~3--8 produced by generate\_performance\_charts.py,\\plot\_combined\_performance\_eos\_intranode.py, and\\plot\_combined\_performance\_eos.py should depict the same generalizable outcomes as in the article, enabling visual verification without reinterpreting raw logs.

\end{description}

\end{document}